\documentclass[journal]{IEEEtran}
\renewcommand{\baselinestretch}{0.982}
\usepackage{amssymb}
\usepackage{amsfonts}
\usepackage{graphicx}
\usepackage{pdfpages}
\usepackage{setspace}
\usepackage{multirow}
\usepackage{amsmath}
\usepackage{epsfig}
\usepackage{array}
\usepackage{color}
\usepackage{subfigure}
\usepackage{xcolor}
\usepackage{amssymb}
\usepackage{url}
\usepackage{mathrsfs}
\usepackage{booktabs}
\usepackage{cite}
\usepackage{bbding}
\usepackage{upgreek}
\usepackage{diagbox, makecell}
\usepackage{amsmath,graphicx,amssymb,delarray,subfigure,verbatim,booktabs}

\ifCLASSINFOpdf
\else
\fi
\definecolor{chcol}{HTML}{EE0000}

\begin{document}
	
%
\title{DBT-Net: Dual-branch federative magnitude and phase estimation with attention-in-attention transformer for monaural speech enhancement}
%
%
%

\author{Guochen Yu,
		Andong~Li, \emph{Student Member, IEEE,}
		Hui Wang,
	    Yutian Wang,
	    Yuxuan Ke,
	     
        and Chengshi~Zheng, \emph{Senior Member, IEEE,}

\thanks{Guochen Yu, Hui Wang, and Yutian Wang are with State Key Laboratory of Media Convergence and Communication, Communication University of China, 100024, Beijing, China. (e-mail: \{yuguochen, hwang, wangyutian\}@cuc.edu.cn)}

\thanks{Andong Li, Yuxuan Ke and Chengshi Zheng are with the Key Laboratory of Noise and Vibration Research, Institute of Acoustics, Chinese Academy of Sciences, Beijing, 100190, China, and also with University of Chinese Academy of Sciences, Beijing, 100049, China. (e-mail: \{liandong, keyuxuan, cszheng\}@mail.ioa.ac.cn)}
\thanks{This work was supported in part by the National Key R\&D Program of China under Grant No. 2021YFF0900700 and in part by the National Natural Science Foundation of China under Grant 61631016.(\textit{Corresponding Author: Chengshi Zheng})}
\thanks{Manuscript received Feb. XX, 2022; revised XXXX XX, XX.}}

%
%

\markboth{Journal of \LaTeX\ Class Files,~Vol.~XX, No.~XX, December~XXXX}%
{Shell \MakeLowercase{\textit{et al.}}: Bare Demo of IEEEtran.cls for Journals}
%



\maketitle

\begin{abstract}
The decoupling-style concept begins to ignite in the speech enhancement area, which decouples the original complex spectrum estimation task into multiple easier sub-tasks {(\emph{i.e.}, magnitude-only recovery and the residual complex spectrum estimation)}, resulting in better performance and easier interpretability. In this paper, we propose a dual-branch federative magnitude and phase estimation framework, dubbed DBT-Net, for monaural speech enhancement, aiming at recovering the coarse- and fine-grained regions of the overall spectrum in parallel. From the complementary perspective, the magnitude estimation branch is designed to filter out dominant noise components in the magnitude domain, while the complex spectrum purification branch is elaborately designed to inpaint the missing spectral details and implicitly estimate the phase information in the complex-valued spectral domain. To facilitate the information flow between each branch, interaction modules are introduced to leverage features learned from one branch, so as to suppress the undesired parts and recover the missing components of the other branch. Instead of adopting the conventional RNNs and temporal convolutional networks for sequence modeling, we employ a novel attention-in-attention transformer-based network within each branch for better feature learning. More specially, it is composed of several adaptive spectro-temporal attention transformer-based modules and an adaptive hierarchical attention module, aiming to capture long-term time-frequency dependencies and further aggregate intermediate hierarchical contextual information. Comprehensive evaluations on the WSJ0-SI84 + DNS-Challenge and VoiceBank + DEMAND dataset demonstrate that the proposed approach consistently outperforms previous advanced systems and yields state-of-the-art performance in terms of speech quality and intelligibility.
\end{abstract}


%
\IEEEpeerreviewmaketitle

\section{Introduction}
Various types of environmental interference may greatly degrade the performance of telecommunication, automatic speech recognition (ASR), and hearing aids in real scenarios. In this regard, monaural speech enhancement (SE) is often necessary, aiming at recovering clean speech from its noise-contaminated mixture to improve speech quality and intelligibility~{\cite{loizou2013speech}}. With the renaissance {of} deep neural networks (DNNs), a plethora of DNN-based approaches have been proposed to ignite the development of SE algorithms for their more powerful capability in suppressing highly non-stationary noise than conventional statistical signal processing-based approaches~{\cite{wang2018supervised}}, particularly under low signal-to-noise ratio (SNR) conditions.

Conventional supervised SE methods based on DNNs usually aim to estimate mask functions or directly predict the spectral magnitude of clean speech in the time-frequency (T-F) domain~{\cite{wang2014training, xu2014regression}}, where the noisy phase remains unaltered when reconstructing the time-domain waveform. This is because that phase information was regarded as unimportant for a long time for the SE task~{\cite{wang1982unimportance}}. Moreover, it is intractable to accurately estimate the phase distribution of clean speech, due to its highly nonstructural characteristic. However, recent studies reported that the unprocessed phase severely degrades the speech perceptual quality especially under low SNR conditions~{\cite{paliwal2011importance}}. To this end, numerous phase-aware SE approaches have been proposed to tackle the phase estimation problem in the time domain or {complex-valued spectral domain.} 
For the former category, the raw waveform is directly used to regenerate enhanced speech without using any T-F representation~{\cite{ defossez2020real, luo2020dual, wang2021tstnn, kinoshita2020improving}}, which diverts around the explicit phase estimation problem. For example, SEGAN~{\cite{pascual2017segan}} proposed a generative adversarial network-based SE method, where a denoising generator directly maps the raw waveform of the clean speech from the mixed raw waveform by adversarial training. {For the latter category, researchers handle the phase estimation in the complex-valued spectral domain~{\cite{williamson2015complex, choi2019phase, yu2021two, tan2019learning}}}, which can be divided into two main streams, namely masking-based and mapping-based. For the first type, a multitude of complex-valued DNN-based algorithms have been proposed to estimate the complex-valued ratio mask (CRM), which can be then applied to real and imaginary (RI) parts of the complex spectrum, so as to recover the magnitude and phase simultaneously~{\cite{williamson2015complex, choi2019phase, yu2021two}}. 
{The second type has leveraged complex spectral mapping networks to directly predict the RI components of the clean complex spectrum~{\cite{tan2019learning}}.} For example, in~{\cite{tan2019learning}}, real-valued convolutional recurrent networks (CRN) were leveraged to directly map the RI components of target speech, where the enhanced RI components were decoded by two decoders respectively. More recently, a handful of multi-stage decoupling-style methods have thrived in the SE area and were demonstrated to achieve a remarkable performance~{\cite{sun2019monaural, li2021two, li2021simultaneous, li2021glance}}. Instead of packing the mapping process into only one black box in the previous single-stage paradigm, these multi-stage methods decoupled the original complex spectrum estimation into optimizing magnitude and phase stage by stage, and alleviated the implicit compensation effect between two targets~{\cite{wang2021compensation}}. Specifically, due to the apparent spectral regularity of magnitude spectra, only the magnitude estimation was involved in the first stage. Subsequently, the complex spectrum refinement was conducted with residual learning in the second stage, which could also implicitly refine the phase. 

Motivated by the aforementioned multi-stage studies, we decompose the complex spectrum estimation in parallel and propose a dual-branch framework involved with a novel transformer-based network, dubbed DBT-Net. From the complementary perspective, DBT-Net takes full advantage of magnitude spectrum-based and complex spectrum-based SE methods to explore the overall spectrum estimation. Specifically, two core branches are elaborately devised in parallel to facilitate the overall spectrum recovery, namely a \textbf{M}agnitude \textbf{E}stimation \textbf{B}ranch (MEB) and an auxiliary \textbf{C}omplex spectrum \textbf{P}urification \textbf{B}ranch (CPB). Due to the apparent spectral regularity of the magnitude spectrum, we seek to construct the filtering system with MEB to coarsely suppress the dominant noise components in the magnitude domain. In parallel, we establish a refining system with CPB to compensate for the lost spectral details and phase mismatch effect in the complex-valued spectral domain. With information interaction between each branch, two branches can flow information and collaboratively facilitate the overall spectrum recovery.

Generally, mainstream SE models leveraged an encoder-decoder structure based on recurrent neural networks (RNN) or convolution neural networks (CNN), which ignored the long-range contextual information during modeling the speech sequences and led to limited denoising performance. Besides, CNN requires more convolutional layers to enlarge the receptive field for model long-term speech sequences, while RNNs suffer from high computational complexity and cannot perform parallel processing. 
Subsequently, convolutional recurrent networks (CRNs)~{\cite{tan2018convolutional}} and temporal convolutional networks (TCNs)~{\cite{pandey2019tcnn}} were proposed for more effective sequence modeling in the SE area, due to their capability of further extracting high-level features and enlarging receptive fields. However, they still lacked sufficient capacity to capture the global contextual information~{\cite{wang2021tstnn, chen2020dual, li2021u, fu2022uformer}}. Additionally, most of them only worked in the time axis, which neglected the correlations among different frequency sub-bands. In this respect, transformer-based approaches have thrived in speech sequence-to-sequence domains for their remarkable performance in capturing the long-term dependency on natural language processing tasks~{\cite{vaswani2017attention}}. In the speech separation and enhancement task, dual-path transformer-based networks were employed for extracting contextual information along both the time and frequency axes~{\cite{wang2021tstnn, chen2020dual, fu2022uformer}}. Nevertheless, they ignored the long-range hierarchical contextual information during sequence modeling, and thus the intermediate feature maps were not fully and effectively exploited. 

To this end, we employ an attention-in-attention transformer network dubbed AIAT within each branch to funnel the global sequence modeling process, which integrates four adaptive time-frequency attention (ATFA) transformers and an adaptive hierarchical attention (AHA) module to form an ``attention-in-attention" (AIA) structure. To be specific, the ATFA transformers can capture the local and global contextual information in the both time and frequency dimension, while the AHA module can flexibly aggregate all the output feature maps of ATFA modules together by a global attention weight. 

{
The major contributions of this paper are summarized as follows.
\begin{itemize}
	\item We propose a dual-branch SE framework to simultaneously recover magnitude and phase information of the clean complex spectrum in parallel, and an attention-in-attention transformer-based network is adopted for sequence modeling. From a complementary perspective, these two core branches can collaboratively obtain the coarse- and fine-grained regions of clean speech, \emph{i.e.}, spectral magnitude and missing details. 
	\item Considering the complementary associations between spectral magnitude and complex spectral details, we introduce information interaction between the two branches, in which the external knowledge is extracted from the magnitude-based branch as assistance to facilitate the residual complex spectral estimation, and vice versa.
	\item Comprehensive experiments on two public corpora show that DBT-Net achieves remarkable results and consistently outperforms the state-of-the-art baselines, while incurring a relatively small model parameter size.
\end{itemize}
}

The remainder of the paper is organized as follows. In Section~{\ref{Section2}}, the target formulation is described in detail. In Section~{\ref{Section3}}, the proposed network architecture is illustrated in detail. The experimental setup is presented in Section~{\ref{Section4}}, while Section~{\ref{Section5}} gives the results and analysis. Finally, some conclusions are drawn in Section~{\ref{Section6}}.
\begin{figure*}[t]
	\centering
	\centerline{\includegraphics[width=1.95\columnwidth]{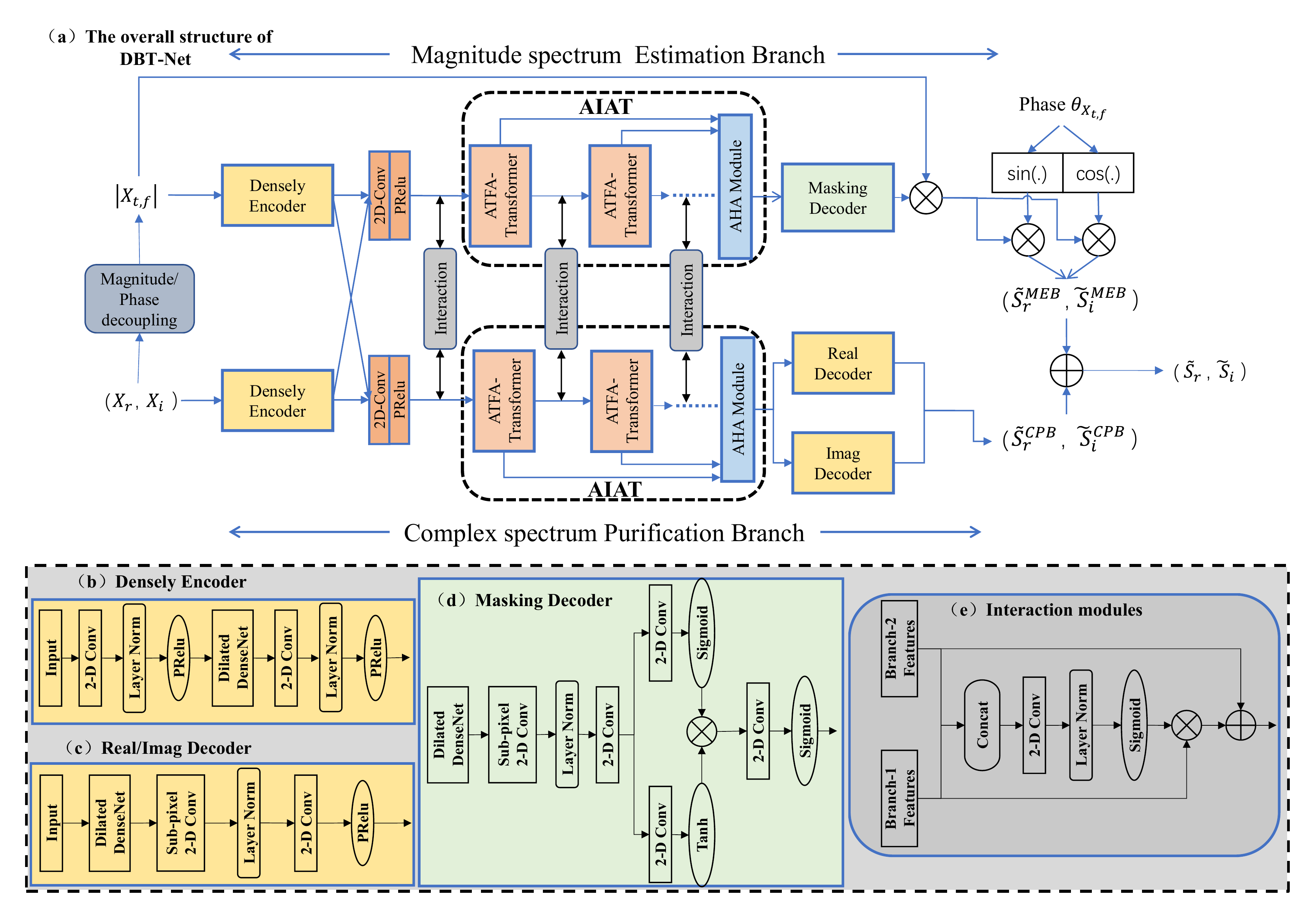}}
	\vspace{-0.6cm}
	\caption{(a) Overall architecture of the proposed DBT-Net. It has two major branches, namely a \textbf{M}agnitude \textbf{E}stimation \textbf{B}ranch (MEB) and an auxiliary \textbf{C}omplex spectrum \textbf{P}urification \textbf{B}ranch (CPB). MEB aims to coarsely suppress the dominant noise components in the magnitude domain, while CPB aims to compensate for the lost spectral details and implicitly estimate the clean phase in the complex-valued spectral domain. (b) The diagram of densely encoder. (c) {The diagram of Real/Imag decoder, where Imag is short for imaginary.}. (d) The diagram of masking decoder. (e) The diagram of interaction modules.}
	\label{fig:diagram-system}
	\vspace{-0.2cm}
\end{figure*}

\vspace{-0.2cm}
\section{Target Formulation\label{Section2}}

\subsection{ Signal model\label{Section21}}

Given a monaural mixture, the noisy signal $x\left [ n \right ]$, clean speech $s\left [ n \right ]$ and noise signals $d\left [ n \right ]$ can be formulated as:

\begin{equation}
	\label{eqn0}
	x\left [ n \right ] =  s\left [ n \right ] + d\left [ n \right ],
\end{equation}
{where $n$ denotes the discrete-time index}. With the short-time Fourier transform (STFT), Eq.~({\ref{eqn0}}) can be transformed into:
\begin{equation}
	\label{eqn2}
	X_{t,f} = S_{t,f} + D_{t,f},
\end{equation}
where $X_{t,f}=\left | X_{t,f} \right |e^{j\theta_{X_{t,f}}} \in \mathbb{C}$, $S_{t,f}=\left | S_{t,f} \right |e^{j\theta_{S_{t,f} }} \in \mathbb{C}$ and $D_{t,f}=\left | D_{t,f} \right |e^{j\theta_{D_{t,f} }}\in \mathbb{C}$ denote the T-F representations of noisy, clean and noise signals in the $(t, f)$ bin index. Note that we omit the time and frequency indices for brevity.
In the Cartesian coordinates, Eq.~({\ref{eqn2}}) can also be written as:
\begin{gather}
	\label{eqn3}
	X =X_{r} + jX_{i} = \left(S_{r}+D_{r}\right) + j\left(S_{i}+D_{i}\right),
\end{gather}
where $X_r = \left | X \right | \cos\left( \theta_{X} \right)$ and $X_i = \left | X \right | \sin\left( \theta_{X} \right)$ denote the real and imaginary parts of the noisy complex spectrum $X$, respectively, with $j^2 = -1$. $S_r$ and $S_i$ have the similar definitions as $X_r$ and $X_i$, as well as $D_r$ and $D_i$.

\vspace{-0.3cm}
\subsection{ Dual-branch strategy\label{Section22}}

The overall diagram of the proposed system is illustrated in Fig.~{\ref{fig:diagram-system}}. It is mainly comprised of two branches, namely a magnitude spectrum estimation branch (MEB) and a complex spectrum purification branch (CPB), which aim at collaboratively estimating the magnitude and phase information of clean speech in parallel. To be specific, we first neglect the intractable phase estimation and only focus on the magnitude estimation. In the MEB path, we feed the magnitude of noisy spectrum $\left | X \right |$ into the network to estimate a gain function $Mask^{MEB}$, which aims at coarsely filtering out dominant noise and recovering the magnitude of the target speech, \emph{i.e.}, $\lvert\widetilde{S}^{MEB}\rvert$. Then the coarsely denoised spectral magnitude is coupled with the corresponding noisy phase $\theta_{X}$ to derive the coarse-estimated RI components of the target spectrum, \emph{i.e.}, $(\widetilde{S}^{MEB}_{r}, \widetilde{S}^{MEB}_{i})$.

As a supplement, we leverage CPB to purify the fine-grained spectral structures which may be lost in the MEB path. That is to say, CPB aims at tackling the residual noise components as well as simultaneously recovering the phase information of the target spectrum. Instead of explicitly estimating the whole complex spectrum, CPB is designed for residual mapping in the complex-valued spectral domain, which can alleviate the overall burden of the network. Finally, we sum the coarse-denoised RI components and the fine-grained complex spectral details together to reconstruct the target complex spectrum. Note that the final output involves both magnitude estimation and complex residual mapping, indicating that the two branches contribute to the target estimation collaboratively. In a nutshell, the whole procedure can be formulated as:
\begin{gather}
	\label{eqn1}
	\lvert\widetilde{S}^{MEB}\rvert = \lvert{X}\rvert \otimes Mask^{MEB},\\
	\widetilde{S}^{MEB}_{r} = \vert\widetilde{S}^{MEB}\rvert \otimes \cos\left( \theta_{X}\right),\\
	\widetilde{S}^{MEB}_{i} = \lvert\widetilde{S}^{MEB}\rvert \otimes \sin\left( \theta_{X} \right), \\
	\widetilde{S}_{r} = \widetilde{S}^{MEB}_{r} + \widetilde{S}^{CPB}_{r}, \\
	\widetilde{S}_{i} = \widetilde{S}^{MEB}_{i} + \widetilde{S}^{CPB}_{i},
\end{gather}
where $\left\{ \widetilde{S}^{CPB}_{r}, \widetilde{S}^{CPB}_{i}  \right\}$ denote the output residual RI components of CPB and $\left\{\widetilde{S}_{r}, \widetilde{S}_{i}\right\}$ denote the final merged estimation of clean RI components. $\otimes$ is the element-wise multiplication operator. The input features of MEB and CPB are denoted as $\lvert{X}\rvert \in \mathbb{R}^{ T\times F\times 1}$ and $X_{com} = Cat(X_r, X_i)\in \mathbb{R}^{T\times F\times 2}$, respectively. Here $T$ is the number of frames and $F$ is the number of frequency bins.

\section{Proposed Architecture\label{Section3}}

\begin{figure}[t]
	\centering
	\centerline{\includegraphics[width=0.95\columnwidth]{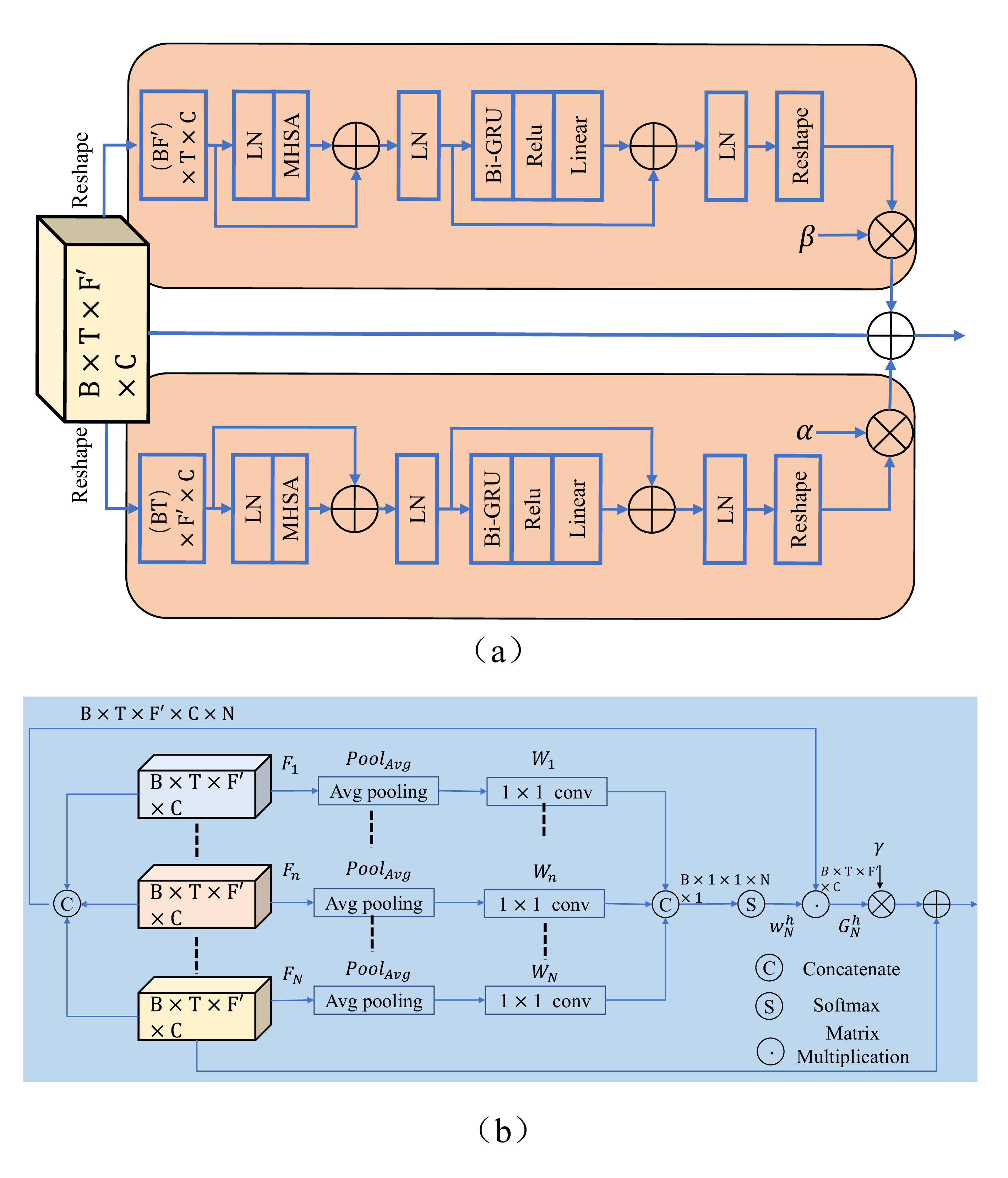}}
	\vspace{-0.4cm}
	\caption{ (a) The diagram of ATFA transformers. (b) The diagram of the AHA module. {“$B$" denotes the batch size.}}
	\vspace{-0.6cm}
	\label{fig:ATFA}	
\end{figure}

In this section, the details of the proposed framework are described. As shown in Fig.~{\ref{fig:diagram-system}} (a), MEB consists of three major components, {namely a densely convolutional encoder}, an AIAT for sequence modeling, and a masking decoder for magnitude spectral gain estimation. {More specifically, the proposed AIAT module consists of four adaptive time-frequency attention transformer-based (ATFAT) modules and an adaptive hierarchical attention (AHA) module, as illustrated in Fig.~{\ref{fig:ATFA}}, where ATFAT aims to capture long-range correlations separately along the temporal and spectral axes and AHA attempts to integrate different intermediate feature maps during sequence modeling.} Note that the output range of the spectral gain is truncated into (0, 1) with the sigmoid activation function.

{The overall topology of CPB is similar to that of MEB, which includes a densely convolutional encoder, an AIAT and two densely decoders}, aiming at estimating the residual real and imaginary parts of the target complex spectrum, respectively. To interact and complement information during sequence modeling, we elaborately design the interaction modules between the two branches, where MEB can better guide the feature learning procedure with the information transformed by CPB, and vice versa. The detailed parameter setup is summarized in Table~{\ref{tab1:parameters}}, while the MEB and CPB share the same configuration except for the number of decoders, where MEB employs only one mask decoder and CPB employs two decoders for the real and imaginary parts of the residual target complex spectrum. 


\renewcommand\arraystretch{1.1}

\begin{table}[htpb]
	\large
	\caption{Detailed parameter setup for the proposed framework. The hyperparameters in the dilated DenseBlock denote the kernel size, strides, the number of kernels and the dilation rate. {“$B$” denotes the batch size ($B$=4 in our experiments). $\delta$ denotes the input dimension, which is set to 1 and 2 for MEB and CPB, respectively.} }
	\vspace{-0.6cm}
	\begin{center}
		\resizebox{\columnwidth}{!}{
			\begin{tabular}{cc|c|c|c}
				\toprule
				&  layer name & input size & hyperparameters & output size\\
				\midrule
				\multirow{6}*{\rotatebox{90}{\textbf{Encoder}}}
				&2-D Conv & $B \times \delta \times T \times 161$ & $1\times 1$, $(1, 1)$, $ 64$ &  $B \times 64\times T\times 161$ \\
				\cline{2-5}
				&DenseBlock\_1 & $B \times64 \times T \times 161$ & $2\times 3$, $(1, 1)$, $64$, $\mathbf{1}$ & $ B \times64 \times T \times 161$\\
				\cline{2-5}
				&DenseBlock\_2 & $B \times64 \times T \times 161$ & $2\times 3$, $(1, 1)$, $64$, $\mathbf{2}$ & $B \times 64 \times T \times 161$\\
				\cline{2-5}
				&DenseBlock\_3 & $B \times64 \times T \times 161$ & $2\times 3$, $(1, 1)$, $64$, $\mathbf{4}$ & $B \times 64 \times T \times 161$\\
				\cline{2-5}
				&DenseBlock\_4 & $B \times64 \times T \times 161$ & $2\times 3$, $(1, 1)$, $64$, $\mathbf{8}$ & $ B \times 64 \times T \times 161$\\
				\cline{2-5}
				&2-D Conv & $B \times64 \times T \times 161$ & $1\times 3$, $(1, 2)$, $64$ & $B \times 64 \times T \times 80$\\
				\midrule
				&2-D Conv (merge)& $B \times 128\times T\times80$ & $1\times 1$, $(1, 1)$, $64$ &$B \times 64 \times T\times80$\\
				\midrule
				\multirow{11}*{\rotatebox{90}{\textbf{ATFA Transformers $\mathbf{\times 4}$}}}
				\multirow{5}*{\rotatebox{90}{\textbf{ATAB}}}
				&reshape & $B \times 64 \times T\times 80$ & -  & $(B\times 80) \times T \times 64$\\
				&MHSA & $(B\times 80) \times T \times 64$ & -  & $(B\times 80) \times T \times 64$\\
				&Bi-GRU & $(B\times 80) \times T \times 64$ &  128 & $(B\times 80) \times T \times 256$\\
				&Linear & $(B\times 80) \times T \times 256$ & 64   & $(B\times 80) \times T \times 64$\\
				&reshape & $(B\times 80) \times T \times 64$ & -  &$B \times 64 \times T \times 80$\\					
				\cline{2-5}
				\multirow{1} *{  }
				\multirow{5}*{\rotatebox{90}{\textbf{AFAB}}} 
				&reshape & $B \times 64 \times T\times 80$ & -  & $(B \times T) \times 80 \times 64 $\\
				&MHSA & $(B\times T) \times 80 \times 64$ & -  & $(B\times T) \times 80 \times 64$\\
				&Bi-GRU & $(B\times T) \times 80 \times 64$ &  128 & $(B\times T) \times 80 \times 256$\\
				&Linear & $(B \times T) \times 80 \times 256$ & 64   & $(B\times T) \times 80 \times 64$\\
				&reshape & $(B\times T) \times 80 \times 64$ & -  & $B \times 64 \times T\times 80$\\
				\cline{2-5}
				&2-D Conv& $B \times 64\times T\times80$ & $1\times 1$, $(1, 1)$, $64$ &$B \times 64 \times T\times80$\\
				\midrule
				& AHA module& $B \times 64 \times T\times 80$ & - & $B \times 64 \times T\times 80$\\
				\midrule					
				\midrule
				\multirow{6}*{\rotatebox{90}{\textbf{\makecell[c] {Real/Imag \\ Decoder}}}}
				&DenseBlock\_1 & $B \times64 \times T \times 80$ & $2\times 3$, $(1, 1)$, $64$, $\mathbf{1}$ & $ B \times64 \times T \times 80$\\
				\cline{2-5}
				&DenseBlock\_2 & $B \times64 \times T \times 80$ & $2\times 3$, $(1, 1)$, $64$, $\mathbf{2}$ & $B \times 64 \times T \times 80$\\
				\cline{2-5}
				&DenseBlock\_3 & $B \times64 \times T \times 80$ & $2\times 3$, $(1, 1)$, $64$, $\mathbf{4}$ & $B \times 64 \times T \times 80$\\
				\cline{2-5}
				&DenseBlock\_4 & $B \times64 \times T \times 80$ & $2\times 3$, $(1, 1)$, $64$, $\mathbf{8}$ & $ B \times 64 \times T \times 80$\\
				\cline{2-5}
				&Sub-pixel Conv & $B \times64 \times T \times 80$ & $1\times 3$, $(1, 2)$, $64$ & $B \times 64 \times T \times 161$\\
				\cline{2-5}
				&2-D Conv & $B \times 64\times T\times 161$ & $1\times 1$, $(1, 1)$, $64$ &$B \times 1 \times T\times161$\\
				\midrule
				\midrule
				\multirow{9}*{\rotatebox{90}{\textbf{\makecell[c] {Masking  Decoder}}}}
				&DenseBlock\_1 & $B \times64 \times T \times 80$ & $2\times 3$, $(1, 1)$, $64$, $\mathbf{1}$ & $ B \times64 \times T \times 80$\\
				\cline{2-5}
				&DenseBlock\_2 & $B \times64 \times T \times 80$ & $2\times 3$, $(1, 1)$, $64$, $\mathbf{2}$ & $B \times 64 \times T \times 80$\\
				\cline{2-5}
				&DenseBlock\_3 & $B \times64 \times T \times 80$ & $2\times 3$, $(1, 1)$, $64$, $\mathbf{4}$ & $B \times 64 \times T \times 80$\\
				\cline{2-5}
				&DenseBlock\_4 & $B \times64 \times T \times 80$ & $2\times 3$, $(1, 1)$, $64$, $\mathbf{8}$ & $ B \times 64 \times T \times 80$\\
				\cline{2-5}
				&Sub-pixel Conv & $B \times64 \times T \times 80$ & $1\times 3$, $(1, 2)$, $64$ & $B \times 64 \times T \times 161$\\
				\cline{2-5}
				&2-D Conv & $B \times 64\times T\times 161$ & $1\times 1$, $(1, 1)$, $1$ &$B \times 1 \times T\times161$\\
				\cline{2-5}
				&2-D Conv\_mask1 & $B \times 1\times T\times 161$ & $1\times 1$, $(1, 1)$, $1$ &$B \times 1 \times T\times161$\\
				&2-D Conv\_mask2 & $B \times 1\times T\times 161$ & $1\times 1$, $(1, 1)$, $1$ &$B \times 1 \times T\times161$\\
				\cline{2-5}
				&2-D Conv & $B \times 1\times T\times 161$ & $1\times 1$, $(1, 1)$, $1$ &$B \times 1 \times T\times161$\\								
				\bottomrule				
		\end{tabular}}
		\label{tab1:parameters}
	\end{center}
	\vspace{-0.3cm}
\end{table}

\subsection{Densely convolutional encoder \label{Section30}}

As illustrated in Fig.~{\ref{fig:diagram-system}} (b), given the input features $\lvert{X}\rvert$ or $X_{com}$, the densely convolutional encoder in each branch is composed of two 2-D convolutional layers, followed by layer normalization (LN) and parametric ReLU (PReLU) activation. {Between these two convolutional layers, a DenseNet ~{\cite{iandola2014densenet}} with four dilated convolutional layers is employed, in which the dilation rates are $\left\{1,2,4,8\right\}$.} The output channel of the first 2-D convolutional layer is set to 64 and keeps unaltered, with kernel size and stride being (1, 1), while the second 2-D convolutional layer halves the dimension of the frequency axis, and sets kernel size and stride to (1, 3) and (1, 2), respectively. The detailed parameter setups for MEB and CPB are presented in Table~{\ref{tab1:parameters}}, where the real/imaginary decoder and masking decoder are employed in CPB and MEB paths, respectively.

\vspace{-0.2cm}

\subsection{ Attention-in-attention transformer \label{Section32}}
{Compared with recurrent neural network (RNN) or convolution neural network (CNN), transformer-based neural network can effectively resolve the long-dependency problem by modeling the speech sequences directly conditioning on context and also operate well in parallel, which has shown remarkable performance in speech enhancement area~{\cite{wang2021tstnn, chen2020dual, li2021u, fu2022uformer}}.
In our attention-in-attention transformer (AIAT), we only utilize the encoder part in the original transformer~{\cite{vaswani2017attention}}, which consists of multi-head scaled dot-product self-attention and position-wise feedforward network similar to~{\cite{wang2021tstnn, chen2020dual}}.}

Before feeding the compressed features into the AIAT in each branch, we concatenate the outputs from two branches in the channel axis and use a 2-D $1 \times 1$ convolution to merge information, followed by PReLU activation. The proposed AIAT module consists of four adaptive time-frequency attention transformer-based (ATFAT) modules and an adaptive hierarchical attention (AHA) module, as illustrated in Fig.~{\ref{fig:ATFA}}. Each ATFAT can strengthen the long-range spectro-temporal dependencies with relatively low computational cost and the AHA module can aggregate different intermediate features to capture global multi-scale contextual information, as pointed out in~{\cite{yu2021cyclegan, yu2021dual,yu2021joint}}. The ATFAT and AHA modules cooperate to form an ``attention-in-attention" structure, which indicates that the output of ATFAT can be further enhanced and integrated by AHA depending on adaptive attention weights.

\subsubsection{ Adaptive time-frequency attention Transformer \label{Section320}} 
To alleviate the heavy computational complexity of conventional self-attention, we introduce an adaptive time-frequency attention (ATFA) mechanism as a lightweight solution to capture long-range correlations exhibited in the temporal and spectral axes, as described in~{\cite{tang2020joint,yu2021cyclegan}}. {As illustrated in Fig.~{\ref{fig:ATFA}}(a), the ATFAT is divided into two sub-branches in the time and frequency axes, namely an adaptive temporal attention branch (ATAB) and an adaptive frequency attention branch (AFAB), which can capture global dependencies along the temporal and spectral dimensions in parallel with two adaptive weights $\alpha$ and $\beta$. In each branch, different from the vanilla transformer, a GRU-based improved transformer~{\cite{wang2021tstnn}} is employed, which is comprised of multi-head self-attention (MHSA) and GRU-based position-wise network, followed by residual connections and LN.} Multi-head self-attention has been widely used in the natural language processing and speech processing areas, because it can leverage the contextual information in the feature maps~{\cite{vaswani2017attention, sperber2018self, tang2020joint, yu2021cyclegan}}. In MHSA modules, the input features are first mapped with different linear projections $h$ times to get queries $(Q)$, keys $(K)$ and values $(V)$ representations, where $h$ indicates the number of heads in MHSA modules. 
Then, the scaled dot-product attention is operated on each head to obtain a weighted sum of the values, where the weight is obtained by an attention function of the query and the corresponding keys. Finally, the attentions of all heads are concatenated and linearly transformed to obtain the final output. Given the input features $F_{in} \in \mathbb{R}^{ B \times T\times F'\times C} $, the attention block in ATAB can be written as: 
\vspace{-0.3cm}

\begin{equation}
	\vspace{-0.1cm}
	\begin{gathered}
		F^{atab} = Reshape(F^{atab}_{in}),\\  
		Q_i = F^{atab}W_{{Q}_i}, K_i = F^{atab}W_{{K}_i},V_i = F^{atab}W_{{V}_i},\\
		head_{i}=Attention({Q}_{i},{K}_{i},{V}_{i})
		=softmax(\frac{Q_{i}(K_{i})^{T}}{\sqrt{C}})V_{i},\\   
		MultiHead\!=\!Concat(head_{1}\!, \!..., \!head_{h})W^{O},\\
		F^{atab}_{mhsa} = LayerNorm(F^{atab}+MultiHead),
	\end{gathered}
\end{equation}
where $F^{atab}\in \mathbb{R}^{(B\times F') \times T\times C}$ denotes the reshaped input of ATAB, $ i \in [1,h]$, and $Q_i,K_i, V_i \in \mathbb{R}^{(B \times F') \times T\times C/h}$ are the linearly mapped queries, keys and values, respectively. $W_{{Q}_i}, W_{{K}_i}, W_{{V}_i}\in{\mathbb{R}^{C\times C/h}}$ and $W^{O}\in{R^{C\times C}}$ are linear transformation matrices. Here, $B$, $T$, $F'$ and $C$ denote the batch size, the frame number, the compressed frequency dimension and the channel number, respectively. {In our model, the number of heads $h$ is set to 4.} Subsequently, inspired by the effectiveness of GRU-based transformer in speech separation and denoising tasks~{\cite{wang2021tstnn, chen2020dual}}, we replace the first fully connected layer of the {feedforward network} in the vanilla transformer with a bi-directional GRU. The final output is calculated by feeding the output from multi-head attention block into GRU-based feedforward network, followed by the residual connections and layer normalization:
\begin{equation}
	\begin{gathered}
		FFN(F^{atab}_{mhsa}) = ReLU(GRU(F^{atab}_{mhsa}))W_{1}+b_{1},\\
		Output=LayerNorm(F^{atab}_{mhsa}+{FFN(F^{atab}_{mhsa})}),\\
		Out_{ATAB} = Reshape(Output),
	\end{gathered}
\end{equation}
where $FFN(\cdot)$ denotes the output of the GRU-based position-wise feedforward network, $W_{1} \in  \mathbb{R}^{C_{ff} \times C}$ denotes the linear transformation, and $b_{1} \in  \mathbb{R}^{C}$ denotes the bias. Here, $C_{ff}=4 \times C$ and $C$ is set to 64 in this module. Then, we reshape the final output of ATAB to the original size, \emph{i.e.}, $Out_{ATAB}\in \mathbb{R}^{B\times T\times F'\times C}$. Analogously, we reshape the compressed input features into $B\times T$ vectors with dimension $ F'\times C$ and feed it into AFAB to calculate the output, \emph{i.e.}, $Out_{AFAB}$, along the frequency axis in parallel. {Finally, the output features of the two branches and the original features are then combined by two learnable adaptive weights $\alpha$ and $\beta$ to obtain the final output of the ATFA module, followed by a PReLU activation and 2-D convolutional layer, which can be formulated as:
\begin{equation}
	\begin{gathered}		
		Out_{ATFA} = F_{in} +\alpha Out_{ATAB} + \beta  Out_{AFAB}
	\end{gathered}
\end{equation}
where $\alpha$ and $\beta$ are initialized to 1 and automatically assigned to suitable values.} After each ATFAT module, a PReLU activation and a 2-D convolutional layer are employed, with kernel size and stride set to (1, 1).

\vspace{-0.3cm}
\subsubsection{ Adaptive hierarchical attention module \label{Section322}}

Given all ATFAT modules' outputs $\left \{ F_m \right \}^{N}_{m=1}, F_m \in \mathbb{R}^{B\times T\times F'\times C}$, the proposed AHA module aims at integrating different intermediate feature maps depending on the global context. Here, $N$ is the number of ATFAT that is set to 4 in the paper. {In the AHA module, we first cascade all the intermediate output of each ATFAT (\emph{i.e.}, $\left \{ F_m \right \}^{N}_{m=1}$) to obtain a global feature map $F^{aha} \in \mathbb{R}^{B\times T\times F'\times C\times N}$. The superscript $aha$ denotes the adaptive hierarchical attention. For each output feature of ATFAT, we employ an average pooling layer $Pool_{Avg}$ and a $1\times1$ convolutional layer to squeeze each ATFAT's output feature into a global representation: $P_m^{aha}=Conv_{1 \times 1}(Pool_{Avg}(F_m))\in \mathbb{R}^{B\times 1\times 1\times 1}$, and then cascade all the pooled outputs as $P^{aha} \in \mathbb{R}^{B\times 1\times 1\times N\times 1}$}. After that, we apply a softmax function to derive the hierarchical attention map $W^{aha} \in \mathbb{R}^{B\times 1\times 1\times N\times 1}$, which can be defined as:
{
\begin{equation}
	\begin{gathered}		
		W^{aha}_m = \frac{\textup{exp}(P_m^{aha})}{\sum_{m=1}^{N}\textup{exp}(P_m^{aha})},
	\end{gathered}
\end{equation} 
where $W^{aha}_m$ denotes the attention weights for the $m$th pooled output with $m \in [1,N]$. Subsequently, we perform the global contextual information modeling by operating a matrix multiplication between $F^{aha}$ and the hierarchical attention weights $W^{aha}$, which can be given by:
{
\begin{equation}
	\begin{gathered}		
		G^{aha} = W^{aha}F^{aha}\\
		=\sum_{m=1}^{N} \frac{\textup{exp}(Conv_{1 \times 1}((Pool_{Avg}(F_m)))}{\sum_{m=1}^{N}\textup{exp}(Conv_{1 \times 1}(Pool_{Avg}(F_m))}F_{m} ,
	\end{gathered}
\end{equation}} 
where $F_m$ denotes the $m$th intermediate output of ATFAT, $Conv_{1 \times 1}$ denotes the $1 \times 1$ convolutional layer and $G^{aha} \in \mathbb{R}^{B\times T\times F'\times C}$ denotes the combined global hierarchical feature.}  {The final output can be obtained by a linear combination of the last ATFAT module output $F_{N}$ and the global contextual feature map $G^{aha}$, \emph{i.e.}, $Out_{AHA}\in \mathbb{R}^{B\times T\times F'\times C}$:
\begin{equation}\\
	\vspace{-2mm}
	\begin{gathered}
		Out_{AHA} = F_{N} + \gamma G^{aha}. \\ 
	\end{gathered}
\end{equation}
where $\gamma$ is a learnable scalar coefficient with zero initialization. During the training process, this adaptive learning weight can automatically learn to assign a suitable value to merge more global contextual information. }
\vspace{-0.2cm}
\subsection{Interaction module \label{Section36}}
{In DBT-Net, the MEB and CPB paths are designed to estimate the spectral magnitude and residual complex spectral details separately, which suggests that these two branches collaboratively facilitate the spectrum restoration. To better guide the sequence modeling process within each branch, we further design an interaction module to exchange information between MEB and CPB. In this way, external information from MEB as assistance can be leveraged to guide CPB to concentrate more on the spectral details which might be lost in the MEB path, and vice versa.}

The structure of the interaction module is shown in Fig.~{\ref{fig:diagram-system}} (e). Taking the interaction of the MEB path as an example, we first concatenate the intermediate feature from MEB (\emph{i.e.}, $ {F}^{MEB}_{in} $) with that from CPB (\emph{i.e.},$ {F}^{CPB}_{in}$). Then, the concatenated feature is fed into the mask module to derive a gain function $\mathcal{G}({{F}^{MEB}_{in},{F}^{CPB}_{in}}) $, which is comprised of 2-D convolution, layer normalization and sigmoid function. To be specific, the gain function automatically learns to filter and preserve different areas of $ {F}^{CPB}_{in} $. A filtered representation is then obtained by multiplying $ \mathcal{G}({{F}^{MEB}_{in},{F}^{CPB}_{in}}) $ with $ {F}^{CPB}_{in} $ elementally. Finally, we add the intermediate feature of the MEB path and the filtered feature of the CPB path to get the final interacted feature of the MEB path, which is subsequently fed into the next sequence modeling block in MEB. {Analogously, we concatenate the intermediate feature from CPB (\emph{i.e.}, $ {F}^{CPB}_{in} $) with that from MEB (\emph{i.e.},$ {F}^{MEB}_{in}$) together, and then feed it into the interaction module to obtain the interacted feature of the CPB path. Note that in Fig. 1 (e), for the interaction operation in the MEB path, “Branch-2 Features" denotes the compressed features from the magnitude branch, and “Branch-1 Features" denotes those from the complex branch, and vice versa in the CPB path.} The whole process is calculated by:  

\begin{equation}
	\begin{aligned} 
		F^{MEB}_{out}&={F}^{MEB}_{in}+{F}^{CPB}_{in}\otimes \mathcal{G}({F}^{MEB}_{in},{F}^{CPB}_{in}),\\
		{F}^{CPB}_{out}&={F}^{CPB}_{in}+{F}^{MEB}_{in}\otimes \mathcal{G}({F}^{CPB}_{in},{F}^{MEB}_{in}),
	\end{aligned}
\end{equation}
where $\mathcal {G}(\cdot)$ denotes the concatenation in the channel axis, convolution, layer normalization and sigmoid operations.
 $\otimes $ denotes element-wise multiplication. In our model, we employ four interaction modules between each ATAFT.

\vspace{-0.2cm}
\subsection{Masking decoder \label{Section34}}

In the MEB path, a masking decoder makes use of the output features from AIAT to obtain the gain function in the magnitude domain for noise suppression. Note that the range of the oracle amplitude mask is considered as unbounded, \emph{i.e.}, $(0, \infty)$, which is intractable to accurately estimate. In this regard, we employ the sigmoid function to scale the value of the spectral gain into $(0, 1)$, and the remaining regions with mask values over 1 can be further compensated by CPB. 

The structure of the masking decoder is presented in Fig.~{\ref{fig:diagram-system}} (d), which mainly consists of a dilated dense block with dilation rates $\left\{1,2,4,8\right\}$, a sub-pixel 2-D convolutional layer with the upsampling factor 2, and a dual-path mask module. {The sub-pixel convolution is utilized to upsample the compressed features, which has demonstrated its effectiveness in both image and speech processing fields~{\cite{shi2016real}}}. Then, a dual-path mask module is performed to obtain the magnitude spectral gain by a 2-D convolution and a dual-path tanh/sigmoid nonlinearity operation, followed by a 2-D convolution and sigmoid activation. The final masked spectral magnitude is obtained by the element-wise multiplication between the input noisy spectral magnitude and the estimated spectral gain. The filtered magnitude in MEB is then coupled with its corresponding noisy phase to obtain {a coarse estimation of the clean complex spectrum}.

\vspace{-0.2cm}
\subsection{Complex decoder \label{Section35}}

In CPB, two decoders are designed to reconstruct the residual RI components in parallel, which aims at refining spectral details in the complex-valued spectral domain. As illustrated in Fig.~{\ref{fig:diagram-system}}(c), both the real and imaginary decoders are composed of a dilated dense block with dilation rates $\left\{1,2,4,8\right\}$, a sub-pixel 2-D convolution, and a $1\times1$ 2-D convolution. {The upsampling factor of the sub-pixel 2-D convolutional layer in the complex decoders is set to 2}, with kernel size (1, 3). The output residual RI components of the CPB path are then incorporated with the coarse-denoised complex spectrum in MEB to obtain the final estimated spectrum. 
\vspace{-0.4cm}

\subsection{Loss function \label{Section37}}

{The loss function of the proposed two-branch model is calculated by the final estimated complex spectrum, which can be expressed as:
\begin{gather}	
	\mathcal{L}^{Mag}=\left \| \sqrt{\left |\widetilde{S}_r  \right |^2+\left |\widetilde{S}_i  \right |^2 } -  \sqrt{\left |S_r  \right |^2+\left |S_i  \right |^2    } \right \|_{F}^2,\\
	\mathcal{L}^{RI}=\left \|\widetilde{S}_r-S_r \right \|_{F}^2 +\left \|\widetilde{S}_i-S_i \right \|_{F}^2,\\
	\mathcal{L}_{Full}=\mu \mathcal{L}^{RI}+(1-\mu ) \mathcal{L}^{Mag}. 
\end{gather}
where $\mathcal{L}^{Mag}$ and $\mathcal{L}^{RI}$ denote the loss functions toward magnitude and RI constraints, respectively. $ \left \|. \right \|_{F}^2$ denotes the mean squared error (MSE) loss.} Here, $\left\{ \widetilde{S}_r, \widetilde{S}_i  \right\}$ represent the RI parts of the estimated speech spectrum, while $\left\{S_r, S_i\right\}$ represent the RI parts of the clean speech spectrum. In Eq. (18), the full loss function is a linear combination of the magnitude and RI loss functions, as it is reported that these two terms together improve the speech quality~{\cite{wang2020complex, li2021two,wisdom2019differentiable}}. With the internal trial, we empirically set $\mu= 0.5$ in the following experiments.

\section{Experiments\label{Section4}}
\label{Sec3}

\subsection{Datasets\label{Section41}}

We first compare the proposed models with several state-of-the-art baselines on a widely used dataset simulated on VoiceBank + DEMAND, and further evaluate our model on the WSJ0-SI84 dataset + DNS challenge.

\textbf{VoiceBank + DEMAND}: The dataset used in this work is publicly available as proposed in~{\cite{valentini2016investigating}}, which is a selection of the VoiceBank corpus~{\cite{veaux2013voice}} with 28 speakers for training and another 2 unseen speakers for testing. The training set includes 11,572 noisy-clean pairs, while the test set contains 824 pairs. For the training set, the audio samples are mixed with one of the 10 noise types, (including two artificial noise processes, \emph{i.e.}, babble and speech shaped noise, and eight real recording noise processes taken from the Demand database~{\cite{thiemann2013diverse}}) at four SNRs, \emph{i.e.}, $\left\{0\rm{dB},5\rm{dB},10\rm{dB},15\rm{dB}\right\}$. The test utterances are created with 5 unseen noise taken from the Demand database at SNRs of $\left\{2.5\rm{dB}, 7.5\rm{dB}, 12.5\rm{dB}, 17.5\rm{dB}\right\}$.

\textbf{WSJ0-SI84 + DNS challenge}: We also investigate the performance of the proposed framework on the WSJ0-SI84 corpus~{\cite{paul1992design}}, which consists of 7138 clean utterances by 83 speakers (42 males and 41 females). We randomly choose 5,428 training utterances and 957 validation utterances from 77 speakers, respectively. In addition, two types of test sets are provided, each of which includes 150 utterances spoken by 6 untrained speakers (3 males and 3 females).
To generate the noisy-clean pairs, we randomly select around 20,000 environmental noises from Interspeech 2020 DNS-Challenge~{\cite{reddy2020interspeech}} as noise set, whose duration is around 55 hours.
During the mixing process, a random noise cut is extracted and then mixed with a randomly sampled utterance under a SNR selected from -5\rm{dB} to 0\rm{dB} with the interval 1dB. As a result, we totally generate around 150,000 and 10,000 noisy-clean pairs for training and validation. The total duration of the training set is around 300 hours. For model evaluation, two challenging untrained noise processes are employed to demonstrate the model generalization capability, namely babble, and factory1 from NOISEX92~{\cite{varga1993assessment}}. Four SNR cases are set, \emph{i.e.}, $\left\{-3\rm{dB}, 0\rm{dB}, 3\rm{dB}, 6\rm{dB}\right\}$, and 150 noisy-clean pairs are generated for each case. 

\subsection{Implementation setup\label{Section42}}

All the utterances are resampled at 16 kHz and chunked to 3 and 4 seconds respectively for {VoiceBank and WSJ0-SI84 datasets}, respectively. The Hanning window of length 20 ms is selected, with 50\% overlap between consecutive frames. The 320-point STFT is utilized and 161-dimension spectral features can be obtained. Due to the efficacy of the compressed magnitude/complex spectrum in {dereverberation and denoising tasks}~{\cite{li2021simultaneous, li2021importance}}, we conduct the power compression toward the magnitude while remaining the phase unaltered, and the optimal compression coefficient is set to 0.5, \emph{i.e.}, $Cat\left(\lvert X \rvert^{0.5}\cos\left({\theta_{X}}\right), \lvert X \rvert^{0.5}\sin\left({\theta_{X}}\right)\right)$ as input, $Cat\left(\lvert S \rvert^{0.5}\cos\left({\theta_{S}}\right), \lvert S \rvert^{0.5}\sin\left({\theta_{S}}\right)\right)$ as target. All the models are optimized using Adam~{\cite{kingma2014adam}} with the learning rate of 8e-4. For VoiceBank + DEMAND benchmark, 80 epochs are conducted for network training in total, while 40 epochs are conducted on WSJ0-SI84 + DNS Challenge benchmark with batch size 4 at the utterance level. {\textbf{We provide enhanced speech samples processed by different models online, and the source code and pretrained model are also released.}}{\footnote{https://github.com/yuguochencuc/DBT-Net} 

\subsection{Baselines\label{Section43}}

In this study, we first compare our model with various advanced systems on WSJ0-SI84 + DNS dataset. For fair comparison, we re-implement all the baselines with non-causal setting, namely BiLSTM~{\cite{chen2016large}}, BiCRN~{\cite{tan2018convolutional}}, GRN~{\cite{tan2018gated}}, DCN~{\cite{pirhosseinloo2019monaural}}, AECNN~{\cite{pandey2019new}}, ConvTasNet~{\cite{luo2019conv}} (Non-causal version), {DPRNN~{\cite{luo2020dual}} (Non-causal version), TSTNN~{\cite{wang2021tstnn}}}, BiDCCRN~{\cite{hu2020dccrn}}, BiGCRN~{\cite{tan2019learning}} and CTS-Net~{\cite{li2021two}} (Non-causal version), {where BiLSTM, BiCRN, DPRNN (Non-causal version) and BiGCRN are similar to LSTM, CRN, original DPRRN and GCRN, respectively, and the only difference is that all the LSTM layers are replaced by their bidirectional versions. Note that for fair comparisons, all the baselines are re-implemented with non-causal configurations.}

BiLSTM and BiCRN are two magnitude-based methods, where the former introduces a bi-directional RNN-based SE model and the latter adopts a typical convolutional recurrent network (CRN) with encoder-decoder architecture. {BiGCRN is an advanced complex spectral mapping network with CRN, where RI components are estimated for magnitude and phase recovery, and all the regular convolutions in the encoder and decoder are replaced by gated linear units (GLUs)~{\cite{dauphin2017language}}}. Both GRN and DCN are based on fully convolutional networks (FCNs), which incorporate dilated GLUs~{\cite{yu2015multi}} and residual connections for magnitude recovery. AECNN is an advanced time-domain model, where the time-domain samples are directly estimated by a typical 1-D U-Net. DPRNN and TSTNN are two dual-path state-of-the-art time-domain methods, where the former employs a dual-path recurrent neural network and the latter employs a dual-path transformer to model the long sequences. {Note that to further improve the performance of DPRNN, we attempt the case of phase-constrained (PCM) loss instead of the original time-domain SI-SNR loss, which was proposed in~{\cite{pandey2021dense}} and demonstrated better performance than SI-SNR loss.} All the baselines are trained with the best parameter configurations mentioned in the reported literature, except that several following modifications are set. Firstly, for BiCRN, except for RI loss, we also introduce the magnitude constraint for better objective performance, which is reported to alleviate the magnitude distortion~{\cite{wang2020complex}}. Secondly, for AECNN, the frame size for input and output is 16384 samples with 50\% frame overlap, \emph{i.e.}, around 1-second contexts can be leveraged for each frame. Besides, in addition to using the reported frequency loss in~{\cite{pandey2019new}}, we also add a time-domain loss as multi-task learning and better performance can be achieved empirically. {Finally, we extend ConvTasNet and DPRNN to the sampling rate of 16 kHz for model comparison, while both of them originally work for the sampling rate of 8 kHz in the speech separation task.}

For the comparison on VoiceBank + DEMAND benchmark, we further adopt several state-of-the-art (SOTA) SE baselines, which includes six time-domain methods (\emph{e.g.}, SEGAN~{\cite{pascual2017segan}}, SERGAN~{\cite{baby2019sergan}}, MHSA-SPK~{\cite{koizumi2020speech}}, TSTNN~{\cite{wang2021tstnn}}, DEMUCS~{\cite{defossez2020real}} and SE-Conformer~{\cite{eesung2021se}}) and ten T-F domain methods(\emph{i.e.}, MMSEGAN~{\cite{soni2018time}}, MericGAN~{\cite{fu2019metricgan}}, DCCRN~{\cite{hu2020dccrn}}, CRGAN~{\cite{zhang2020loss}}, RDL-Net~{\cite{nikzad2020deep}}, T-GSA~{\cite{kim2020t}}, PHASEN~{\cite{yin2020phasen}}, GaGNet~{\cite{li2021glance}} and MetricGAN+~{\cite{fu2021metricgan+}}). SEGAN, SERGAN, MMSEGAN, MetricGAN, CRGAN and MetricGAN+ are all based on generative adversarial networks (GANs), in which a generator ($G$) aims to conduct the enhancement process and a discriminator ($D$) aims to distinguish between the generated speech features and the real clean ones. Note that MetricGAN and MetricGAN+ optimize the generator with respect to one or multiple evaluation metrics such as PESQ and STOI by a pretrained metric-related discriminator. {MHSA-SPK, T-GSA, TSTNN and SE-Conformer all employ multi-head self-attention mechanisms to capture long-term temporal sequence information for better performance}, where the latter three models are conducted on transformer-based networks. RDL-Net introduces a novel residual-dense lattice network incorporated into a Deep Xi-MMSE-LSA based framework~{\cite{nicolson2019deep, zhang2020deepmmse}} to estimate \textit{a priori} SNR. DEMUCS is a SOTA real-time SE model working on the raw waveform domain, which introduces the time-domain L1 loss together with a multi-resolution STFT loss over the spectral magnitude. PHASEN and GaGNet both belong to advanced two-branch phase-aware SE methods, where both magnitude and phase are recovered simultaneously.

\renewcommand\arraystretch{1.2}
\begin{table*}[t]
	\caption{Objective results toward ablation study. Due to space limit, we only report the average values at $\left\{-3\rm{dB},0\rm{dB},3\rm{dB}\right\}$ in terms of PESQ, ESTOI and SDR for factory1 noise. {“Feat." denotes the input feature types and “Inter." denotes whether to use the interaction module. “Para." represents the number of trainable parameters. ``MACs" and “TBT" denote the multiply-accumulate operations per second and training batch time, respectively. “D" denotes the downsampling operations in the densely encoder.}}
	\Huge
	\centering
	\resizebox{0.96\textwidth}{!}{
		\begin{tabular}{cc|c|c|ccc|c|ccc|cccc|cccc|cccc}
			\toprule
			&Metrics &\multirow{2}*{id} &\multirow{2}*{Feat.} &\multicolumn{3}{c|}{AIA structure}  &\multirow{2}*{Inter.} &Para. &{MACs} &{TBT}
			&\multicolumn{4}{c|}{PESQ} &\multicolumn{4}{c|}{ESTOI(\%)} &\multicolumn{4}{c}{SDR(dB)} \\
			\cline{1-2} \cline{5-7} \cline{12-23}
			&SNR(dB) &  & & ATAB & AFAB  & AHA  &  & (M) & {(G/s)} & {(s)} &-3 &0 &3 &\multicolumn{1}{c|}{Avg.}  &-3 &0 &3  &\multicolumn{1}{c|}{Avg.}  &-3 &0 &3  &\multicolumn{1}{c}{Avg.}\\
			\cline{1-23}
			&\multicolumn{1}{c|}{{Unprocessed}} &- &- &- &- &- &- &- &- &- &1.58 &1.76 &1.97 &\multicolumn{1}{c|}{1.77}  &33.37 &42.43 &52.18 &\multicolumn{1}{c|}{42.66}  &-2.92 &0.04 &3.04  &\multicolumn{1}{c}{0.16}\\
			\cline{1-23}
			\multicolumn{22}{c}{\textbf{Magnitude-branch models}} \\ \hline
			&\multicolumn{1}{c|}{MEB-Net } &1  &Mag &\Checkmark &\XSolidBrush &\XSolidBrush &\multicolumn{1}{c|}{-}  &\textbf{0.64} &{8.04} &{0.60}  &2.32 &2.48 &2.72 &\multicolumn{1}{c|}{2.51}  &60.17 &66.32 &72.21 &\multicolumn{1}{c|}{66.23} &4.58 &7.12 &9.67 &\multicolumn{1}{c}{7.12}\\
			&\multicolumn{1}{c|}{MEB-Net} &2  &Mag &\XSolidBrush &\Checkmark &\XSolidBrush &\multicolumn{1}{c|}{-}  & \textbf{0.64} &{7.99} &{0.53} &2.38 &2.57 &2.78 &\multicolumn{1}{c|}{2.58}  &62.01 &69.56 &76.09 &\multicolumn{1}{c|}{69.22} &5.78 &8.03 &10.56 &\multicolumn{1}{c}{8.12}\\
			&\multicolumn{1}{c|}{MEB-Net} &3  &Mag &\Checkmark &\Checkmark &\XSolidBrush &\multicolumn{1}{c|}{-}  & 0.90 &{9.71} &{1.35} &2.47 &2.71 &2.92 &\multicolumn{1}{c|}{2.70}  &63.91 &72.01 &77.89 &\multicolumn{1}{c|}{71.27} &6.42 &8.86 &11.02 &\multicolumn{1}{c}{8.77}\\
			&\multicolumn{1}{c|}{MEB-Net} &4  &Mag &\Checkmark &\Checkmark &\Checkmark &\multicolumn{1}{c|}{-}  &0.90 &{9.72} &{1.38}  &2.52 &2.76 &3.00 &\multicolumn{1}{c|}{2.76}  &64.48 &72.24 &78.55 &\multicolumn{1}{c|}{71.76} &6.99 &9.05 &11.14 &\multicolumn{1}{c}{9.06}\\
			
			\midrule
			\multicolumn{22}{c}{\textbf{Complex-branch models}} \\ \hline
			&\multicolumn{1}{c|}{CPB-Net} &1  &RI &\Checkmark &\XSolidBrush &\XSolidBrush &\multicolumn{1}{c|}{-}  &0.91 &{10.22} &{0.66} &2.48 &2.74 &3.01 &\multicolumn{1}{c|}{2.74}  &64.27 &71.16 &78.01 &\multicolumn{1}{c|}{71.14} &7.03 &9.68 &11.02 &\multicolumn{1}{c}{9.24}\\
			
			&\multicolumn{1}{c|}{CPB-Net } &2  &RI &\XSolidBrush &\Checkmark &\XSolidBrush &\multicolumn{1}{c|}{-}  &0.91 &{10.17} &{0.58} &2.54 &2.81 &3.06 &\multicolumn{1}{c|}{2.80}  &66.16 &73.68 &79.51 &\multicolumn{1}{c|}{73.12} &7.79 &10.03 &11.61 &\multicolumn{1}{c}{9.81}\\
			
			&\multicolumn{1}{c|}{CPB-Net } &3  &RI &\Checkmark &\Checkmark &\XSolidBrush &\multicolumn{1}{c|}{-}  & 1.18 &{11.89} &{1.43} &2.63 &2.89 &3.12 &\multicolumn{1}{c|}{2.88}  &68.23 &75.46 &81.03 &\multicolumn{1}{c|}{74.91} &8.47 &10.82 &12.04 &\multicolumn{1}{c}{10.44}\\			
			&\multicolumn{1}{c|}{CPB-Net } &4  &RI &\Checkmark &\Checkmark &\Checkmark &\multicolumn{1}{c|}{-}  &1.18 &{11.89} &{1.47} &2.70 &2.97 &3.18 &\multicolumn{1}{c|}{2.95}  &69.56 &76.78 &82.22 &\multicolumn{1}{c|}{76.19} &9.34 &11.21 &12.79 &\multicolumn{1}{c}{11.11}\\			
			
			\midrule
			\multicolumn{22}{c}{\textbf{Dual-branch models}} \\ \hline
			&\multicolumn{1}{c|}{{DCB-Net}} &{1}  &{RI + RI} &{\Checkmark} &{\Checkmark} &{\Checkmark} &\multicolumn{1}{c|}{{\Checkmark}}  &{3.18} &{42.76} &{2.27} &{2.74} &{3.02} &{3.22} &\multicolumn{1}{c|}{{2.99}}  &{70.28} &{76.93} &{82.69} &\multicolumn{1}{c|}{{76.63}} &{9.56} &{11.71} &{13.49} &\multicolumn{1}{c}{{11.59}}\\
				
			&\multicolumn{1}{c|}{{DBT-Net$^{\spadesuit}$}} &{1}  &{Mag + RI} &{\Checkmark} &{\Checkmark} &{\Checkmark} &\multicolumn{1}{c|}{{\Checkmark}}  &{2.91} &{40.59} &{2.20} &{2.83} &{3.06} &{3.27} &\multicolumn{1}{c|}{{3.05}}  &{72.60} &{79.43} &{84.47} &\multicolumn{1}{c|}{{78.83}} &{9.97} &{11.99} &{13.85} &\multicolumn{1}{c}{{11.94}}\\
			&\multicolumn{1}{c|}{{DBT-Net (D=2)}} &{1}  &{Mag + RI} &{\Checkmark} &{\Checkmark} &{\Checkmark} &\multicolumn{1}{c|}{{\Checkmark}}  &{2.98} &{23.65} &{1.54} &{2.73} &{2.98} &{3.19} &\multicolumn{1}{c|}{{2.96}}  &{71.34} &{78.84} &{84.02} &\multicolumn{1}{c|}{{78.07}} &{9.29} &{11.64} &{13.62} &\multicolumn{1}{c}{{11.52}}\\
			
			&\multicolumn{1}{c|}{{DBT-Net (D=3)}} &{2}  &{Mag + RI} &{\Checkmark} &{\Checkmark} &{\Checkmark} &\multicolumn{1}{c|}{{\Checkmark}}  &{3.08} &{12.48} &{0.96} &{2.65} &{2.86} &{3.12} &\multicolumn{1}{c|}{{2.88}}  &{69.01} &{75.94} &{81.51} &\multicolumn{1}{c|}{{76.25}} &{8.59} &{10.47} &{12.71} &\multicolumn{1}{c}{{10.76}}\\
			
			&\multicolumn{1}{c|}{{DBT-Net (D=4)}} &{3}  &{Mag + RI} &{\Checkmark} &{\Checkmark} &{\Checkmark} &\multicolumn{1}{c|}{{\Checkmark}}  &{3.18} &{6.92} &{0.62} &{2.57} &{2.79} &{3.03} &\multicolumn{1}{c|}{{2.79}}  &{67.82} &{74.65} &{80.78} &\multicolumn{1}{c|}{{74.42}} &{8.13} &{10.21} &{12.17} &\multicolumn{1}{c}{{10.17}}\\
			\midrule
			
			&\multicolumn{1}{c|}{DBT-Net} &1  &Mag + RI &\Checkmark &\XSolidBrush &\XSolidBrush &\multicolumn{1}{c|}{\XSolidBrush}  &2.08 &{27.67} &{1.46}  &2.62 &2.90 &3.09 &\multicolumn{1}{c|}{2.87}  &68.28 &76.32 &80.99 &\multicolumn{1}{c|}{75.20} &8.62 &10.93 &12.19 &\multicolumn{1}{c}{10.58}\\
			&\multicolumn{1}{c|}{DBT-Net} &2  &Mag + RI &\XSolidBrush &\Checkmark &\XSolidBrush &\multicolumn{1}{c|}{\XSolidBrush}  & 2.08 &{27.49} &{1.29} &2.66 &2.92 &3.11 &\multicolumn{1}{c|}{2.90}  &70.19 &76.83 &82.04 &\multicolumn{1}{c|}{76.35} &9.31 &11.12 &12.85 &\multicolumn{1}{c}{11.09}\\
			
			&\multicolumn{1}{c|}{DBT-Net} &3  &Mag + RI &\Checkmark &\Checkmark &\XSolidBrush &\multicolumn{1}{c|}{\XSolidBrush}  &2.80 &{40.12} &{2.06} &2.74 &3.01 &3.16 &\multicolumn{1}{c|}{2.97}  &71.16 &77.69 &83.87 &\multicolumn{1}{c|}{77.57} &9.53 &11.65 &13.42 &\multicolumn{1}{c}{11.53}\\	
			
			&\multicolumn{1}{c|}{DBT-Net} &4  &Mag + RI &\Checkmark &\Checkmark &\Checkmark &\multicolumn{1}{c|}{\XSolidBrush}  &2.81 &{40.13} &{2.13} &2.87 &3.10 &3.29 &\multicolumn{1}{c|}{3.09} &74.32 &80.56 &85.09 &\multicolumn{1}{c|}{79.99} &10.49 &12.37 &14.06 &\multicolumn{1}{c}{12.31}\\
			&\multicolumn{1}{c|}{DBT-Net} &5  &Mag + RI &\Checkmark &\Checkmark &\Checkmark &\multicolumn{1}{c|}{\Checkmark}  &2.91&{40.59} &{2.19}  &\textbf{2.89} &\textbf{3.13} &\textbf{3.32} &\multicolumn{1}{c|}{\textbf{3.11}}  &\textbf{75.07} &\textbf{81.11} &\textbf{85.55} &\multicolumn{1}{c|}{\textbf{80.57}} &\textbf{10.60} &\textbf{12.57} &\textbf{14.36} &\multicolumn{1}{c}{\textbf{12.51}}\\
			\midrule

			\midrule
			
			\bottomrule
	\end{tabular}}
	\label{tbl:ablation study}
\end{table*}

\subsection{Evaluation metrics\label{Section44}} 
In the evaluation on WSJ0-SI84 + DNS benchmark, we use perceptual evaluation of speech quality (PESQ)~{\cite{rix2001perceptual}}, extended short-time objective intelligibility (ESTOI)~{\cite{jensen2016algorithm}} and SDR~{\cite{vincent2007first}} as the objective metrics to evaluate the enhancement performance of different models. PESQ is used to evaluate perceptual speech quality, whose score ranges from $-0.5$ to $4.5$. Note that we use the narrow-band version recommended in ITU-T P.862.2 for WSJ0-SI84 + DNS dataset. ESTOI is the extended version of STOI to measure speech intelligibility~{\cite{taal2010short}}, where the mutual independence assumption among frequency bands is canceled. The ESTOI score ranges from 0 to 1. SDR is widely used in blind speech separation and evaluates the level of speech distortion in the waveform. For VoiceBank + DEMAND corpus, we use wide-band PESQ (WB-PESQ), STOI, and three MOS metrics~{\cite{hu2007evaluation}} (\emph{i.e.,} CSIG, CBAK, and COVL) to evaluate the speech quality. Here, CSIG, CBAK and COVL are designed to measure signal distortion, the background noise quality and the overall audio quality evaluation, respectively. All three MOS scores range from 1 to 5. {Besides the aforementioned intrusive metrics, DNSMOS is also adopted to evaluate the perceptual speech quality~{\cite{reddy2020dnsmos}},} which is a robust non-intrusive perceptual speech quality metric serving as a proxy for subjective scores, ranging from 1 to 5. Higher values of all aforementioned metrics indicate better speech quality. 

\section{Results and Analysis\label{Section5}}
\label{Sec4}
 
\subsection{Ablation study on WSJ0-SI84 + DNS Challenge \label{Section50}}
For the ablation study, we create a smaller dataset on WSJ0-SI84 + DNS Challenge. For training, we establish a 15000, 1500 noisy-clean pairs also at the SNR range of $\left\{-5\rm{dB}, -4\rm{dB}, -3\rm{dB}, -2\rm{dB}, -1\rm{dB}, 0\rm{dB}\right\}$ for training and validation, respectively. The total duration for the training set is about 30 hours. For testing, three SNR cases are set with factory1 noise from NOISEX92 on both seen and unseen speakers, \emph{i.e.}, $\left\{-3\rm{dB}, 0\rm{dB}, 3\rm{dB}\right\}$, and 150 noisy-clean pairs are generated for each case. The results of the ablation study are presented in Table~{\ref{tbl:ablation study}}, which studies the impacts of different AIA structures, the dual-branch strategy and the interaction modules. To be specific, MEB-Net (1)-(4) only estimate the spectral gain in the magnitude domain and retains the phase information unaltered, while CPB-Net (1)-(4) estimate the RI parts of the clean complex spectrum. {
The configurations of MEB-Net and CPB-Net are similar to those presented in Table~{\ref{tab1:parameters}}. MEB-Net adopts a densely convolutional encoder, an AIAT without information interaction, and a magnitude masking decoder. CPB-Net adopts a similar encoder, an AIAT, and two separate decoders to decode RI components of the clean complex spectrum. By merging MEB-Net and CPB-Net, DBT-Net (1)-(5) aim at estimating the magnitude spectral gain and the residual RI components of the clean complex spectrum in parallel. Additionally, another two dual-branch models are implemented to investigate the effectiveness of separately estimating the spectral magnitude and residual complex spectral details. Specifically, we employ a dual-branch CPB-Net, dubbed \textbf{DCB-Net}, to estimate the RI components and the residual complex spectral details, which can be formulated as:
\begin{gather}
	\widetilde{S}_{r}^{DCB} = \widetilde{S}^{CB^1}_{r} + \widetilde{S}^{CB^2}_{r}, \\
	\widetilde{S}_{i}^{DCB} = \widetilde{S}^{CB^1}_{i} + \widetilde{S}^{CB^2}_{i},
\end{gather}
where $\left\{ \widetilde{S}^{CB^1}_{r},\widetilde{S}^{CB^1}_{i}  \right\}$ denote the estimated RI components by the first branch in DCB-Net and $\left\{ \widetilde{S}^{CB^2}_{r},\widetilde{S}^{CB^2}_{i}  \right\}$ denote the estimated residual RI components by the second branch. $\left\{\widetilde{S}_{r}^{DCB} , \widetilde{S}_{i}^{DCB} \right\}$ denote the final RI estimation by DCB-Net.
Then, we implement another dual-branch network, dubbed \textbf{DBT-Net$^{\spadesuit}$}, to estimate the spectral magnitude and the whole RI components of the clean complex spectrum instead of the residual RI components in parallel. In DBT-Net$^{\spadesuit}$, the magnitude branch (\emph{i.e.}, MEB$^{\spadesuit}$) aims at estimating the spectral magnitude while the complex branch (\emph{i.e.}, CPB$^{\spadesuit}$) aims at restoring the whole RI components of the clean complex spectrum instead of estimating the residual complex spectrum as in our proposed reconstruction strategy. Finally, we average the spectral magnitude estimated by MEB$^{\spadesuit}$ and CPB$^{\spadesuit}$ and use the estimated phase by CPB$^{\spadesuit}$ to obtain the final RI components of the clean complex spectrum. The whole procedure can be formulated as:
\begin{gather}
	\lvert\widetilde{S}^{MEB^{\spadesuit}}\rvert = \lvert{X}\rvert \otimes Mask^{MEB^{\spadesuit}},\\
	\lvert\widetilde{S}^{CPB^{\spadesuit}}\rvert = \sqrt{\left |\widetilde{S}_r^{CPB^{\spadesuit}}  \right |^2+\left |\widetilde{S}_i^{CPB^{\spadesuit}}  \right |^2 },\\	
	\lvert\widetilde{S}^{{\spadesuit}}\rvert = (\lvert\widetilde{S}^{MEB^{\spadesuit}}\rvert + \lvert\widetilde{S}^{CPB^{\spadesuit}}\rvert)/2, \\
	\theta_{\widetilde{S}^{{\spadesuit}}} =  \arctan \left(|\widetilde{S}_r^{CPB^{\spadesuit}} / |\widetilde{S}_i^{CPB^{\spadesuit}}\right)
\end{gather}
where $\left\{ \widetilde{S}_r^{CPB^{\spadesuit}}, \widetilde{S}_i^{CPB^{\spadesuit}} \right\}$ denote the output RI components of the CPB$^{\spadesuit}$ path. $\lvert\widetilde{S}^{MEB^{\spadesuit}}\rvert$ and $\lvert\widetilde{S}^{CPB^{\spadesuit}}\rvert$ represent the estimated spectral magnitude of MEB$^{\spadesuit}$ and CPB$^{\spadesuit}$. $\lvert\widetilde{S}^{{\spadesuit}}\rvert$ and $\theta_{\widetilde{S}^{\spadesuit}}$ denote the final output spectral magnitude and phase of the clean complex spectrum, respectively.
}

We also investigate different attention mechanisms in single-branch and dual-branch approaches. For example, MEB-Net (1) and (2) only adopt the adaptive temporal attention branch (ATAB) or the adaptive temporal attention branch (AFAB) in the proposed transformer-based network, while MEB-Net (3) utilizes the combination of ATAB and ATFB as the ATFA modules. Then, we merge the ATFA modules and the AHA module as the attention-in-attention structure in MEB-Net (4). CPB-Net (1)-(4) and DBT-Net (1)-(4) utilize the same attention mechanisms as in MEB-Net (1)-(4), respectively. Finally, we add the interaction modules in DBT-Net (5) to investigate the impact of information interaction. {Moreover, we implement another three dual-branch frameworks similar to DBT-Net (5), namely DBT-Net (D=2), DBT-Net (D=3) and DBT-Net (D=4), to investigate the impact of using more downsampling layers in our model. Specifically, DBT-Net (D=2) utilizes two 2-D convolutional downsampling layers in the encoder and two symmetrical 2-D convolutional upsampling layers in the decoder, which is set to 1 in DBT-Net (1)-(5). That is to say, the frequency dimension of the encoded features is downsampled to 40 in DBT-Net (D=2). Analogously, DBT-Net (D=3) and DBT-Net (D=4) employ three and four 2-D convolutional downsampling layers in the encoders, respectively.} The values are averaged upon both seen and unseen speaker conditions. 

\subsubsection{Effect of AIA structure}
We first analyze the effect of different attention mechanisms in our sequence modeling network. As shown in Table~{\ref{tbl:ablation study}}, taking the magnitude-based methods (MEB-Net) as an example, the objective performance of the enhanced speech is severely limited. When combining ATAB and AFAB to capture both spectro-temporal dependencies, MEB-Net (3) dramatically outperforms MEB-Net (1) and MEB-Net (2) in terms of PESQ, ESTOI and SDR. For example, from MEB-Net (1) to MEB-Net (3), around 0.19, 5.04\% and 1.65\rm{dB} score improvements are obtained for PESQ, ESTOI and SDR, averaging under seen and unseen speaker conditions. Similar results are also observed in the complex-path models (CPB-Net(1)-(3)) and dual-branch models (DBT-Net (1)-(3)). This indicates the merit of simultaneously capturing spectro-temporal contextual information in parallel. Then, by adding the adaptive hierarchical attention (AHA) module as an attention-in-attention topology, consistently better speech performance can be obtained in all metrics. For example, MEB-Net (4) yields average 0.06 PESQ, 0.49\% ESTOI and 0.29\rm{dB} SDR improvements over MEB-Net (3) with nearly same parameter burden. A similar tendency is also observed for CPB-Net and DBT-Net. This verifies the effectiveness of the proposed AIA structure in improving speech quality and intelligibility.
\subsubsection{Effect of dual-branch strategy}
We then investigate the impact of the dual-branch strategy, \emph{i.e.}, magnitude estimation path and complex refining path. First, when using the single branch, the complex spectrum mapping approaches, \emph{i.e.}, CPB-Nets, consistently surpass the magnitude spectrum estimate approaches, i.e, MEB-Nets. This indicates the importance of phase recovery in improving speech quality and intelligibility. However, although phase recovery is involved, the performance of single-branch methods is limited. The latent reason is that when directly estimating the whole RI components, the phase is implicitly optimized and the magnitude may deviate from its optimal optimization path, leading to sub-optimal solutions. When federatively merging the magnitude estimation branch and complex refining branch, DBT-Nets dramatically outperform the single-branch approaches. For example, DBT-Net (4) provides average 0.21, 3.80\% and 1.87\rm{dB} score improvements over CPB-Net (4) for PESQ, ESTOI and SDR, averaging under seen and unseen speaker conditions. {Subsequently, when compared with other dual-branch models (\emph{i.e.}, DCB-Net and DBT-Net$^{\spadesuit}$), the proposed dual-branch model achieves consistently better performance. When only estimating the RI components of the clean complex spectrum directly, DCB-Net obtains relatively marginal improvements over the single-branch CPB-Nets, which indicates the effectiveness of decoupling the magnitude and phase recovery in two heads. Meanwhile, DBT-Net (5) considerably surpasses DBT-Net$^{\spadesuit}$ in terms of all metrics, while suffering a similar model computational cost. This verifies the superiority of the dual-branch strategy, \emph{i.e.}, eliminating the dominant noise in the magnitude domain and restoring the residual complex details by two branches.}

{In addition, we provide the segmental SNR improvements ($\Delta$SegSNR)~{\cite{mermelstein1979evaluation}} of the single-branch (MEB-Net (4) and CPB-Net (4)) and dual-branch methods (\emph{i.e.}, DBT-Net (5)) over the unprocessed mixtures in both seen and unseen speaker conditions, as shown in Fig.~{\ref{fig:SSNR}} (a) and (b)}. One can observe that DBT-Net produces significantly larger SegSNR improvements than the single-branch baselines, with which a more than average 10\rm{dB} $\Delta$SegSNR improvement is achieved at -3\rm{dB}, 0\rm{dB}, 3\rm{dB} SNR cases. These observations fully emphasize the remarkable superiority of the proposed dual-branch strategy in speech quality and intelligibility.

\subsubsection{Effect of interaction modules}
We finally investigate the effect of the information interaction modules between two branches. After introducing the information flow between the MEB path and the CPB path, One can observe that DBT-Net (5) consistently surpasses DBT-Net (4) with only a little increased parameter size. For example, from DBT-Net (4) to DBT-Net (5), around 0.02, 0.58\% and 0.20\rm{dB} score improvements are obtained for PESQ, ESTOI and SDR, averaging under seen and unseen speaker conditions. These results show that the interaction modules indeed facilitate the simultaneous magnitude estimation and complex spectral details refinement, resulting in better spectrum restoration.

\vspace{-0.2cm}
{
\subsubsection{Model Complexity Discussion}
In Table~{\ref{tbl:ablation study}}, we provide detailed model complexity comparisons in terms of the number of parameters, the multiply-accumulate operations (MACs) per second, and the training batch time (TBT) among models. Specifically, we measure the MACs using an utterance with a duration of one second, and the TBT is evaluated with one-second utterances and the batch size of 4 on Tesla M40 with 24 GB of RAM. From Table~{\ref{tbl:ablation study}}, we can observe that although the proposed DBT-Net achieves low trainable parameters, the MACs are relatively high, leading to a large computational cost. This is because, to better model the dependencies along the different frequency bands by adaptive frequency attention branch (AFAB), we only employ one downsampling layer along with the frequency axis, leading to relatively high frequency-dimension encoded features and large MACs.}
 
{Then we investigate the trade-off between the computational cost and objective performance improvements. As shown in Table~{\ref{tbl:ablation study}}, when using more downsampling operations in the encoder, DBT-Net (D=2), (D=3) and (D=4) can effectively reduce the MACs and TBT with a slightly increased parameter size. However, due to the insufficient sequence modeling in the frequency axis caused by the decreased frequency dimension of features, DBT-Net (D=2)-(D=4) also decrease the speech enhancement performance with the increasing number of downsampling layers along the frequency axis. For example, DBT-Net (D=2) decreases around 0.15 PESQ, 1.51\% ETOI and 0.99\rm{dB} SDR scores by DBT-Net (5), while decreasing around 16.94 G/s MACs. This indicates that utilizing more downsampling layers can reduce the computation cost with somewhat increased parameter burden, and meanwhile, it decreases the speech enhancement performance. In practical applications, we can balance parameter number and computation cost flexibly by adjusting the number of downsampling layers and convolutional kernels according to different equipment requirements. In the following experiments, to achieve the best performance with a small model size, all the models only employ the downsampling-upsampling operation once in the densely encoder-decoder, and the frequency dimension is set to 80 during the sequence modeling.  For practical applications, it is often desired to reduce the parameter burden and computational complexity, which can be studied in future research. }

\vspace{-0.4cm}
\begin{figure}
	\centering
	\centerline{\includegraphics[width=1\columnwidth]{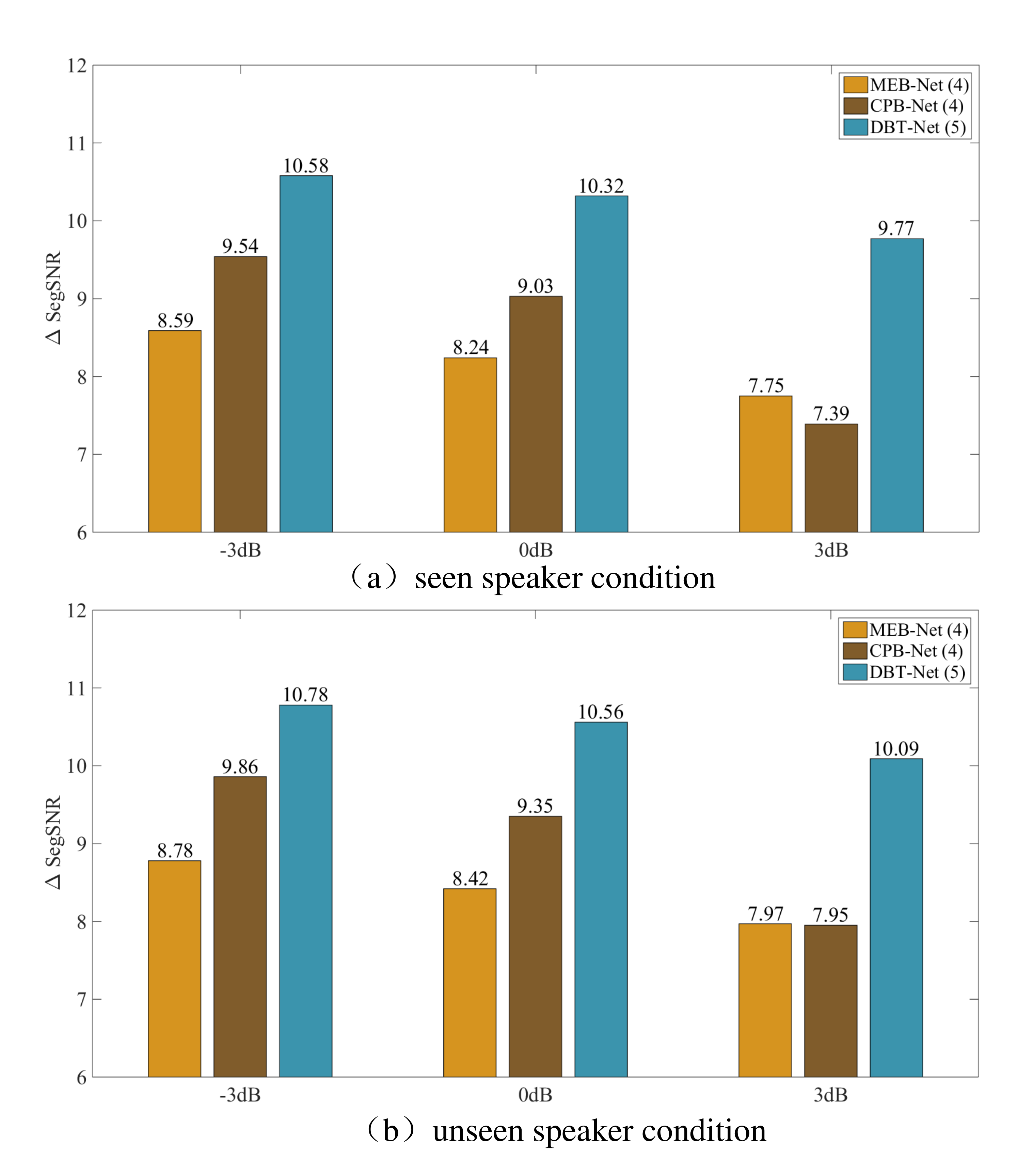}}
	\vspace{-0.4cm}
	\caption{Segmental SNR improvements ($\Delta$SegSNR) over the unpropossed mixtures for $-3$, $0$ and $3\rm{dB}$ under seen and unseen speaker conditions. }
	\label{fig:SSNR}
	\vspace{-0.2cm}	
\end{figure}

\renewcommand\arraystretch{0.9}
\begin{table*}[t]
	\caption{Objective result comparisons among different models in terms of PESQ, ESTOI and SDR for babble and factory1 noises in the seen speaker test set. {“Cau." denotes whether to use the causal setup.}}
	\Huge
	\centering
	\resizebox{0.9\textwidth}{!}{
		\begin{tabular}{cc|c|c|ccccc|ccccc|ccccc}
			\toprule
			&Metrics &\multirow{2}*{\rotatebox{90}{Cau.}}&\multirow{2}*{Feat.}
			&\multicolumn{5}{c|}{PESQ} &\multicolumn{5}{c|}{ESTOI(\%)} &\multicolumn{5}{c}{SDR(dB)} \\
			\cline{1-2}\cline{5-19}
			&SNR(dB) &  & &-3 &0 &3 &6 &\multicolumn{1}{c|}{Avg.}  &-3 &0 &3  &6 &\multicolumn{1}{c|}{Avg.}  &-3 &0 &3 &6 &\multicolumn{1}{c}{Avg.}\\
			\cline{1-19}
			\multirow{8}*{\rotatebox{90}{Babble Noise}}
			&\multicolumn{1}{|c|}{Noisy} &- &- &1.72 &1.88 &2.06 &2.25 &\multicolumn{1}{c|}{1.98}  &34.76 &43.27 &51.95 &61.61 &\multicolumn{1}{c|}{47.90}  &-2.94 &0.04 &3.03 &6.03 &\multicolumn{1}{c}{1.54}\\
			&\multicolumn{1}{|c|}{BiLSTM~{\cite{chen2016large}}} &\XSolidBrush &Mag  &2.41 &2.68 &2.90 &3.09 &\multicolumn{1}{c|}{2.77}  &67.20 &75.22 &80.26 &84.39 &\multicolumn{1}{c|}{76.77}  &4.86 &7.33 &9.53 &11.66 &\multicolumn{1}{c}{8.35}\\
			&\multicolumn{1}{|c|}{BiCRN~{\cite{tan2018convolutional}}} &\XSolidBrush &Mag &2.38 &2.64 &2.84 &3.04 &\multicolumn{1}{c|}{2.73}  &66.05 &73.67 &79.10 &83.63 &\multicolumn{1}{c|}{75.61}  &4.90 &7.36 &9.52 &11.78 &\multicolumn{1}{c}{8.39}\\
			&\multicolumn{1}{|c|}{GRN~{\cite{tan2018gated}}} &\XSolidBrush &Mag  &2.16 &2.37 &2.54 &2.70 &\multicolumn{1}{c|}{2.44}  &59.92 &68.20 &74.35 &79.74 &\multicolumn{1}{c|}{70.55}  &4.37 &6.82 &9.05 &11.25 &\multicolumn{1}{c}{7.87}\\
			&\multicolumn{1}{|c|}{DCN~{\cite{pirhosseinloo2019monaural}}} &\XSolidBrush &Mag &2.15 &2.41 &2.62 &2.82 &\multicolumn{1}{c|}{2.50} &57.38 &66.26 &72.99 &78.59 &\multicolumn{1}{c|}{68.81}  &4.16 &6.75 &9.03 &11.21 &\multicolumn{1}{c}{7.79}\\
			&\multicolumn{1}{|c|}{AECNN~{\cite{pandey2019new}}} &\XSolidBrush  &waveform &2.30 &2.63 &2.90 &3.13 &\multicolumn{1}{c|}{2.74}  &64.84 &73.49 &79.62 &84.14 &\multicolumn{1}{c|}{75.52}  &7.16 &9.84 &11.96 &13.95 &\multicolumn{1}{c}{10.73}\\
			&\multicolumn{1}{|c|}{ConvTasNet~{\cite{luo2019conv}}} &\XSolidBrush &Waveform &2.67 &2.96 &3.18 &3.35 &\multicolumn{1}{c|}{3.04} &74.97 &81.67 &85.81 &88.82 &\multicolumn{1}{c|}{82.82} &9.51 &12.21 &14.20 &15.98 &\multicolumn{1}{c}{12.96}\\
			
			&\multicolumn{1}{|c|}{{DPRNN~{\cite{luo2020dual}}}} &{\XSolidBrush} &{Waveform} &{2.86} &{3.14} &{3.33} &{3.47} &\multicolumn{1}{c|}{{3.20}} &{78.09} &{81.67} &{86.25} &{88.33} &\multicolumn{1}{c|}{{83.59}} &{10.09} &{12.21} &{14.04} &{15.98} &\multicolumn{1}{c}{{13.08}}\\
			
			&\multicolumn{1}{|c|}{{TSTNN~{\cite{wang2021tstnn}}}} &{\XSolidBrush} &{Waveform} &{2.62} &{2.92} &{3.17} &{3.38} &\multicolumn{1}{c|}{{3.02}} &{72.23} &{74.99} &{83.62} &{88.01} &\multicolumn{1}{c|}{{79.71}} &{9.01} &{12.18} &{13.48} &{15.21} &\multicolumn{1}{c}{{12.37}}\\
			&\multicolumn{1}{|c|}{BiGCRN~{\cite{tan2019learning}}} &\XSolidBrush &RI  &2.59 &2.88 &3.11 &3.28 &\multicolumn{1}{c|}{2.97}  &70.12 &77.90 &83.09 &86.58 &\multicolumn{1}{c|}{79.42}  &7.62 &10.14 &12.23 &14.17 &\multicolumn{1}{c}{11.04}\\
			
			&\multicolumn{1}{|c|}{BiDCCRN~{\cite{hu2020dccrn}}} &\XSolidBrush &RI &2.46 &2.77 &2.99 &3.22 &\multicolumn{1}{c|}{2.86}  &65.62 &73.04 &79.81 &86.06 &\multicolumn{1}{c|}{76.13}  &7.36 &9.78 &12.21 &14.39 &\multicolumn{1}{c}{10.93}\\
			&\multicolumn{1}{|c|}{CTS-Net~{\cite{li2021two}}} &\XSolidBrush &Mag+RI &2.76 &3.05 &3.24 &3.39 &\multicolumn{1}{c|}{3.11}  &74.94 &81.33 &85.22 &88.23 &\multicolumn{1}{c|}{82.43}  &9.16 &11.61 &13.46 &15.14 &\multicolumn{1}{c}{12.34}\\			
			
			&\multicolumn{1}{|c|}{MEB-Net(Pro.)} &\XSolidBrush &Mag &2.53 &2.82 &3.06 &3.27 &\multicolumn{1}{c|}{2.92}  &67.93 &75.56 &81.36 &86.19 &\multicolumn{1}{c|}{77.76}  &7.24 &8.86 &11.28 &13.26 &\multicolumn{1}{c}{10.16}\\
			
			&\multicolumn{1}{|c|}{CPB-Net(Pro.)} &\XSolidBrush &RI &2.69 &2.98 &3.17 &3.39 &\multicolumn{1}{c|}{3.06}  &72.79 &75.95 &84.37 &88.49 &\multicolumn{1}{c|}{80.40}  &8.81 &11.37 &12.88 &14.69 &\multicolumn{1}{c}{11.94}\\
			
			&\multicolumn{1}{|c|}{DBT-Net(Pro.)} &\XSolidBrush &Mag+RI  &\textbf{2.89} &\textbf{3.18} &\textbf{3.38} &\textbf{3.55} &\multicolumn{1}{c|}{\textbf{3.25}}  &\textbf{78.22}  &\textbf{83.78} &\textbf{87.51} &\textbf{90.01} &\multicolumn{1}{c|}{\textbf{84.06}}  &\textbf{10.57} &\textbf{12.72} &\textbf{14.50} &\textbf{16.20} &\multicolumn{1}{c}{\textbf{13.50}}\\
			
			\midrule
			\multirow{1}*{\rotatebox{90}{Factory1 Noise}}
			&\multicolumn{1}{|c|}{Noisy} &- &Mag  &1.60 &1.78 &1.99 &2.20 &\multicolumn{1}{c|}{1.89}  &34.76 &43.73 &54.04 &63.90
			&\multicolumn{1}{c|}{49.11}  &-2.92 &0.05 &3.04 &6.03 &\multicolumn{1}{c}{1.55}\\
			&\multicolumn{1}{|c|}{BiLSTM~{\cite{chen2016large}}} &\XSolidBrush &Mag  &2.53 &2.75 &2.96 &3.14 &\multicolumn{1}{c|}{2.85} &68.40 &75.18 &80.54 &84.78 &\multicolumn{1}{c|}{77.23}  &5.89 &8.03 &10.20 &12.19 &\multicolumn{1}{c}{9.08}\\
			&\multicolumn{1}{|c|}{BiCRN~{\cite{tan2018convolutional}}} &\XSolidBrush &Mag  &2.48 &2.70 &2.90 &3.07 &\multicolumn{1}{c|}{2.79}  &66.10 &73.16 &79.12 &83.87 &\multicolumn{1}{c|}{75.56} &5.73 &7.96 &10.21 &12.31 &\multicolumn{1}{c}{9.05}\\
			&\multicolumn{1}{|c|}{GRN~{\cite{tan2018gated}}} &\XSolidBrush &Mag &2.26 &2.43 &2.58 &2.70 &\multicolumn{1}{c|}{2.49}  &62.06 &69.70 &76.09 &81.25 &\multicolumn{1}{c|}{72.28} &5.60 &7.86 &10.07 &12.10 &\multicolumn{1}{c}{8.91}\\
			&\multicolumn{1}{|c|}{DCN~{\cite{pirhosseinloo2019monaural}}} &\XSolidBrush &Mag &2.32 &2.55 &2.74 &2.92 &\multicolumn{1}{c|}{2.63}&59.47 &68.05 &74.67 &80.18 &\multicolumn{1}{c|}{70.59} &5.60 &7.85 &10.02 &12.03 &\multicolumn{1}{c}{8.88}\\
			&\multicolumn{1}{|c|}{AECNN~{\cite{pandey2019new}}} &\XSolidBrush &Waveform &2.44 &2.72 &2.97 &3.16 &\multicolumn{1}{c|}{2.82} &64.12 &72.74 &78.86 &83.54 &\multicolumn{1}{c|}{74.82} &8.14 &10.26 &12.14 &13.92 &\multicolumn{1}{c}{11.11}\\
			&\multicolumn{1}{|c|}{ConvTasNet~{\cite{luo2019conv}}} &\XSolidBrush &Waveform &2.79 &3.02 &3.20 &3.36 &\multicolumn{1}{c|}{3.09} &75.11 &81.04 &85.20 &88.39 &\multicolumn{1}{c|}{82.43} &10.16 &12.14 &13.96 &15.64 &\multicolumn{1}{c}{12.97}\\
			&\multicolumn{1}{|c|}{{DPRNN~{\cite{luo2020dual}}}} &{\XSolidBrush} &{Waveform} &{2.91} &{3.13} &{3.30} &{3.43} &\multicolumn{1}{c|}{{3.19}} &{77.03} &{81.85} &{86.21} &{89.15} &\multicolumn{1}{c|}{{83.56}} &{10.46} &{12.72} &{14.08} &{15.74} &\multicolumn{1}{c}{{13.25}}\\
			
			&\multicolumn{1}{|c|}{{TSTNN~{\cite{wang2021tstnn}}}} &{\XSolidBrush} &{Waveform} &{2.72} &{2.94} &{3.21} &{3.42} &\multicolumn{1}{c|}{{3.08}} &{71.46} &{76.23} &{82.62} &{87.82} &\multicolumn{1}{c|}{{79.53}} &{9.82} &{11.67} &{13.75} &{15.03} &\multicolumn{1}{c}{{12.57}}\\
						
			&\multicolumn{1}{|c|}{BiGCRN~{\cite{tan2019learning}}} &\XSolidBrush &RI &2.69 &2.95 &3.16 &3.31 &\multicolumn{1}{c|}{3.02} &69.24 &77.11 &82.63 &86.45 &\multicolumn{1}{c|}{78.86} &7.78 &10.20 &12.21 &14.05 &\multicolumn{1}{c}{11.06}\\
			
			&\multicolumn{1}{|c|}{BiDCCRN~{\cite{hu2020dccrn}}} &\XSolidBrush &RI &2.49 &2.82 &3.07 &3.26 &\multicolumn{1}{c|}{2.91}  &65.07 &74.37 &81.25 &87.02 &\multicolumn{1}{c|}{76.93}  &8.01 &10.71 &12.43 &14.37 &\multicolumn{1}{c}{11.38}\\
			&\multicolumn{1}{|c|}{CTS-Net~{\cite{li2021two}}} &\XSolidBrush &Mag+RI &2.80 &3.03 &3.21 &3.36 &\multicolumn{1}{c|}{3.10}  &73.90 &80.02 &84.27 &87.60 &\multicolumn{1}{c|}{81.45}  &9.60 &11.56 &13.33 &14.99 &\multicolumn{1}{c}{12.37}\\	
			&\multicolumn{1}{|c|}{MEB-Net(Pro.)} &\XSolidBrush &Mag &2.63 &2.88 &3.11 
			&3.32  &\multicolumn{1}{c|}{2.98}  &67.06 &74.82 &80.58 &85.76 &\multicolumn{1}{c|}{77.06}  &7.31 &9.29 &11.36 &13.46 &\multicolumn{1}{c}{10.35}\\
			
			&\multicolumn{1}{|c|}{CPB-Net(Pro.)} &\XSolidBrush &RI &2.74 &3.00 &3.21 &3.41 &\multicolumn{1}{c|}{3.09}  &71.86 &76.67 &83.54 &88.01 &\multicolumn{1}{c|}{80.02}  &9.73 &11.64 &13.50 &14.99 &\multicolumn{1}{c}{12.47}\\
			
			&\multicolumn{1}{|c|}{DBT-Net(Pro.)} &\XSolidBrush &Mag+RI  &\textbf{2.92} &\textbf{3.15} &\textbf{3.33} &\textbf{3.49} &\multicolumn{1}{c|}{\textbf{3.22}}  &\textbf{77.11}  &\textbf{82.62} &\textbf{86.48} &\textbf{89.69} &\multicolumn{1}{c|}{\textbf{83.98}}  &\textbf{10.68} &\textbf{12.57} &\textbf{14.28} &\textbf{15.99} &\multicolumn{1}{c}{\textbf{13.38}}\\			
			\bottomrule
	\end{tabular}}
	\label{tbl:seen-babble-factory1-objective}
	
\end{table*}

\renewcommand\arraystretch{0.9}
\begin{table*}[t]
	\caption{Objective result comparisons among different models in terms of PESQ, ESTOI and SDR for babble and factory1 noises in the unseen speaker test set.}
	\label{tbl:unseen-babble-factory1-objective}
	\Huge
	\centering
	\resizebox{0.9\textwidth}{!}{
		\begin{tabular}{cc|c|c|ccccc|ccccc|ccccc}
			\toprule
			&Metrics &\multirow{2}*{\rotatebox{90}{Cau.}}&\multirow{2}*{Feat.}
			&\multicolumn{5}{c|}{PESQ} &\multicolumn{5}{c|}{ESTOI(\%)} &\multicolumn{5}{c}{SDR(dB)} \\
			\cline{1-2}\cline{5-19}
			&SNR(dB) & &  &-3 &0 &3 &6 &\multicolumn{1}{c|}{Avg.}  &-3 &0 &3  &6 &\multicolumn{1}{c|}{Avg.}  &-3 &0 &3 &6 &\multicolumn{1}{c}{Avg.}\\
			\cline{1-19}
			\multirow{8}*{\rotatebox{90}{Babble Noise}}
			&\multicolumn{1}{|c|}{Noisy} &- &- &1.64 &1.82 &2.01 &2.23 &\multicolumn{1}{c|}{1.93}  &31.51 &39.66 &48.21 &57.74 &\multicolumn{1}{c|}{44.28}  &-2.93 &0.05 &3.03 &6.03 &\multicolumn{1}{c}{1.55}\\
			&\multicolumn{1}{|c|}{BiLSTM~{\cite{chen2016large}}} &\XSolidBrush &Mag  &2.30 &2.58 &2.79 &2.99 &\multicolumn{1}{c|}{2.67}  &63.26 &71.51 &77.17 &85.02 &\multicolumn{1}{c|}{74.24}  &4.92 &7.29 &9.41 &11.48 &\multicolumn{1}{c}{8.28}\\
			&\multicolumn{1}{|c|}{BiCRN~{\cite{tan2018convolutional}}} &\XSolidBrush &Mag &2.25 &2.55 &2.77 &2.97 &\multicolumn{1}{c|}{2.64} &61.51 &70.71 &76.73 &81.98 &\multicolumn{1}{c|}{72.73}  &4.73 &7.42 &9.76 &12.08 &\multicolumn{1}{c}{8.50}\\
			&\multicolumn{1}{|c|}{GRN~{\cite{tan2018gated}}} &\XSolidBrush &Mag &2.08 &2.32 &2.54 &2.72 &\multicolumn{1}{c|}{2.42}  &56.52 &65.57 &72.66 &78.85 &\multicolumn{1}{c|}{68.39}  &4.21 &6.76 &9.18 &11.62 &\multicolumn{1}{c}{7.94}\\
			&\multicolumn{1}{|c|}{DCN~{\cite{pirhosseinloo2019monaural}}} &\XSolidBrush &Mag &2.03 &2.32 &2.57 &2.79 &\multicolumn{1}{c|}{2.43}  &53.19 &63.17 &70.81 &77.49 &\multicolumn{1}{c|}{66.17}  &3.92 &6.71 &9.14 &11.56 &\multicolumn{1}{c}{7.83}\\
			&\multicolumn{1}{|c|}{AECNN~{\cite{pandey2019new}}} &\XSolidBrush &Waveform  &2.24 &2.59 &2.86 &3.10 &\multicolumn{1}{c|}{2.69}  &62.64 &72.11 &78.37 &83.57 &\multicolumn{1}{c|}{74.18}  &6.26 &9.90 &12.12 &14.20 &\multicolumn{1}{c}{10.62}\\
			&\multicolumn{1}{|c|}{ConvTasNet~{\cite{luo2019conv}}} &\XSolidBrush &Waveform &2.58 &2.89 &3.12 &3.30 &\multicolumn{1}{c|}{2.97} &72.36 &79.81 &84.49 &87.94 &\multicolumn{1}{c|}{81.14} &9.50 &12.00 &14.08 &16.00 &\multicolumn{1}{c}{12.89}\\
			
			&\multicolumn{1}{|c|}{{DPRNN~{\cite{luo2020dual}}}} &{\XSolidBrush} &{Waveform} &{2.82} &{3.07} &{3.31} &{3.43} &\multicolumn{1}{c|}{{3.16}} &{76.86} &{82.45} &{86.51} &{87.94} &\multicolumn{1}{c|}{{83.44}} &{10.19} &{12.74} &{14.13} &{15.86} &\multicolumn{1}{c}{{13.23}}\\

			&\multicolumn{1}{|c|}{{TSTNN~{\cite{wang2021tstnn}}}} &{\XSolidBrush} &{Waveform} &{2.59} &{2.93} &{3.12} &{3.36} &\multicolumn{1}{c|}{{3.00}} &{71.97} &{74.64} &{82.37} &{87.95} &\multicolumn{1}{c|}{{79.23}} &{9.98} &{12.63} &{14.03} &{15.79} &\multicolumn{1}{c}{{13.11}}\\
			
			&\multicolumn{1}{|c|}{BiGCRN~{\cite{tan2019learning}}} &\XSolidBrush  &RI &2.55 &2.85 &3.08 &3.27 &\multicolumn{1}{c|}{2.94}  &68.41 &76.68 &82.16 &86.38 &\multicolumn{1}{c|}{78.41}  &7.77 &10.42 &12.55 &14.60 &\multicolumn{1}{c}{11.34}\\
			
			&\multicolumn{1}{|c|}{BiDCCRN~{\cite{hu2020dccrn}}} &\XSolidBrush  &RI &2.41 &2.71 &2.96 &3.19 &\multicolumn{1}{c|}{2.82}  &63.47 &71.94 &77.93 &84.92 &\multicolumn{1}{c|}{74.56}  &7.31 &9.86 &12.37 &14.55 &\multicolumn{1}{c}{11.02}\\
			&\multicolumn{1}{|c|}{CTS-Net~{\cite{li2021two}}} &\XSolidBrush  &Mag+RI &2.68 &3.00 &3.22 &3.39 &\multicolumn{1}{c|}{3.07}  &73.11 &80.10 &84.49 &87.76 &\multicolumn{1}{c|}{81.37}  &9.34 &11.74 &13.70 &15.51 &\multicolumn{1}{c}{12.57}\\	
			&\multicolumn{1}{|c|}{MEB-Net(Pro.)} &\XSolidBrush &Mag &2.50 &2.78 &3.02 &3.25 &\multicolumn{1}{c|}{2.89}  &66.90 &74.42 &80.60 &86.37 &\multicolumn{1}{c|}{77.07}  &7.31 &9.32 &11.42 &13.59 &\multicolumn{1}{c}{10.41}\\
			
			&\multicolumn{1}{|c|}{CPB-Net(Pro.)} &\XSolidBrush  &RI &2.62 &2.97 &3.21 &3.41 &\multicolumn{1}{c|}{3.05}  &72.47 &75.15 &83.85 &88.34 &\multicolumn{1}{c|}{79.95}  &9.16 &11.83 &13.64 &15.24 &\multicolumn{1}{c}{12.47}\\
			
			&\multicolumn{1}{|c|}{DBT-Net(Pro.)} &\XSolidBrush  &Mag+RI &\textbf{2.91} &\textbf{3.18} &\textbf{3.39} &\textbf{3.57} &\multicolumn{1}{c|}{\textbf{3.26}}  &\textbf{77.74}  &\textbf{83.47} &\textbf{87.37} &\textbf{90.13} &\multicolumn{1}{c|}{\textbf{84.68}}  &\textbf{11.09} &\textbf{13.23} &\textbf{15.01} &\textbf{16.81} &\multicolumn{1}{c}{\textbf{14.03}}\\
			
			\midrule
			\multirow{1}*{\rotatebox{90}{Factory1 Noise}}
			&\multicolumn{1}{|c|}{Noisy} &- &-  &1.55 &1.75 &1.96 &2.17 &\multicolumn{1}{c|}{1.86}  &31.97 &41.13 &50.32 &59.78
			&\multicolumn{1}{c|}{45.80}  &-2.92 &0.04 &3.04 &6.03 &\multicolumn{1}{c}{1.55}\\
			&\multicolumn{1}{|c|}{BiLSTM~{\cite{chen2016large}}} &\XSolidBrush &Mag &2.43 &2.65 &2.86 &3.04 &\multicolumn{1}{c|}{2.75}  &64.51 &71.81 &77.68 &82.01 &\multicolumn{1}{c|}{74.00}  &5.96 &8.04 &10.03 &11.91 &\multicolumn{1}{c}{8.99}\\
			&\multicolumn{1}{|c|}{BiCRN~{\cite{tan2018convolutional}}} &\XSolidBrush  &Mag  &2.40 &2.63 &2.84 &3.01 &\multicolumn{1}{c|}{2.72}  &62.72 &70.76 &77.02 &81.81 &\multicolumn{1}{c|}{73.08}  &5.90 &8.20 &10.39 &12.57 &\multicolumn{1}{c}{9.26}\\
			&\multicolumn{1}{|c|}{GRN~{\cite{tan2018gated}}} &\XSolidBrush &Mag  &2.24 &2.44 &2.61 &2.75 &\multicolumn{1}{c|}{2.51}  &59.86 &68.38 &74.89 &80.01 &\multicolumn{1}{c|}{70.79}  &5.78 &8.08 &10.31 &12.46 &\multicolumn{1}{c}{9.16}\\
			&\multicolumn{1}{|c|}{DCN~{\cite{pirhosseinloo2019monaural}}} &\XSolidBrush &Mag &2.26 &2.51 &2.73 &2.90 &\multicolumn{1}{c|}{2.60}  &56.89 &66.11 &73.40 &78.94 &\multicolumn{1}{c|}{68.84}  &5.67 &8.05 &10.31 &12.42 &\multicolumn{1}{c}{9.11}\\
			&\multicolumn{1}{|c|}{AECNN~{\cite{pandey2019new}}} &\XSolidBrush &Waveform  &2.40 &2.69 &2.94 &3.13 &\multicolumn{1}{c|}{2.79}  &62.09 &71.23 &78.00 &82.54 &\multicolumn{1}{c|}{73.47}  &8.30 &10.42 &12.35 &14.14 &\multicolumn{1}{c}{11.30}\\
			&\multicolumn{1}{|c|}{ConvTasNet~{\cite{luo2019conv}}} &\XSolidBrush &Waveform &2.73 &2.98 &3.17 &3.33 &\multicolumn{1}{c|}{3.06} &73.12 &79.68 &84.39 &87.42 &\multicolumn{1}{c|}{81.15} &10.32 &12.31 &14.04 &15.72 &\multicolumn{1}{c}{13.10}\\
			
			&\multicolumn{1}{|c|}{{DPRNN~{\cite{luo2020dual}}}} &{\XSolidBrush} &{Waveform} &{2.85} &{3.11} &{3.17} &{3.38} &\multicolumn{1}{c|}{{3.13}} &{75.92} &{81.53} &{85.63} &{88.32} &\multicolumn{1}{c|}{{82.85}} &{10.84} &{12.89} &{14.32} &{15.85} &\multicolumn{1}{c}{{13.47}}\\
			
			&\multicolumn{1}{|c|}{{TSTNN~{\cite{wang2021tstnn}}}} &{\XSolidBrush} &{Waveform} &{2.70} &{2.94} &{3.21} &{3.40} &\multicolumn{1}{c|}{{3.08}} &{71.01} &{76.18} &{83.05} &{87.29} &\multicolumn{1}{c|}{{79.38}} &{10.01} &{12.07} &{13.80} &{15.27} &\multicolumn{1}{c}{{12.81}}\\
			&\multicolumn{1}{|c|}{BiGCRN~{\cite{tan2019learning}}} &\XSolidBrush  &RI  &2.65 &2.93 &3.14 &3.30 &\multicolumn{1}{c|}{3.01}  &66.94 &76.06 &81.90 &85.77 &\multicolumn{1}{c|}{77.67} &8.19 &10.54 &12.61 &14.42 &\multicolumn{1}{c}{11.44}\\
			
			&\multicolumn{1}{|c|}{BiDCCRN~{\cite{hu2020dccrn}}} &\XSolidBrush  &RI &2.45 &2.78 &3.04 &3.23 &\multicolumn{1}{c|}{2.88}  &64.54 &73.87 &80.81 &86.38 &\multicolumn{1}{c|}{76.39}  &8.37 &10.92 &12.81 &14.98 &\multicolumn{1}{c}{11.77}\\
			&\multicolumn{1}{|c|}{CTS-Net~{\cite{li2021two}}} &\XSolidBrush &Mag+RI  &2.78 &3.02 &3.22 &3.37 &\multicolumn{1}{c|}{3.09}  &72.72 &78.33 &83.95 &87.07 &\multicolumn{1}{c|}{80.52}  &9.92 &11.92 &13.68 &15.35 &\multicolumn{1}{c}{12.72}\\	
			&\multicolumn{1}{|c|}{MEB-Net(Pro.)} &\XSolidBrush  &Mag &2.61 &2.85 &3.08 &3.26 &\multicolumn{1}{c|}{2.95}  &65.48 &74.41 &79.22 &86.68 &\multicolumn{1}{c|}{76.45}  &7.76 &9.61 &11.91 &14.16 &\multicolumn{1}{c}{10.86}\\
			
			&\multicolumn{1}{|c|}{CPB-Net(Pro.)} &\XSolidBrush  &RI &2.75 &3.00 &3.22 &3.42 &\multicolumn{1}{c|}{3.10}  &71.17 &78.50 &84.11 &88.05 &\multicolumn{1}{c|}{80.46}  &9.94 &11.98 &13.81 &15.25 &\multicolumn{1}{c}{12.74}\\
			
			&\multicolumn{1}{|c|}{DBT-Net(Pro.)} &\XSolidBrush &Mag+RI  &\textbf{2.97} &\textbf{3.19} &\textbf{3.37} &\textbf{3.51} &\multicolumn{1}{c|}{\textbf{3.26}}  &\textbf{77.01}  &\textbf{82.62} &\textbf{86.64} &\textbf{89.53} &\multicolumn{1}{c|}{\textbf{83.95}}  &\textbf{11.35} &\textbf{13.20} &\textbf{14.88} &\textbf{16.51} &\multicolumn{1}{c}{\textbf{13.99}}\\			
			\bottomrule
	\end{tabular}}
\end{table*}

\subsection{Performance comparison with baselines using WSJ0-SI84 + DNS Challenge dataset\label{Section52}}

{Based on previous ablation studies, DBT-Net (5) is selected as the default configuration of the proposed framework. Besides, we also evaluate the two proposed single-branch methods with their best performance as shown in Table~{\ref{tbl:ablation study}} (\emph{i.e.}, MEB-Net (4) and CPB-Net (4)) as the reference.} Then we compare the performance of the proposed methods with advanced non-causal time and T-F domain baselines in terms of PESQ, ESTOI and SDR, whose objective results are presented in Tables~{\ref{tbl:seen-babble-factory1-objective}} and ~{\ref{tbl:unseen-babble-factory1-objective}}. {From the results in Tables~{\ref{tbl:seen-babble-factory1-objective}} and~{\ref{tbl:unseen-babble-factory1-objective}}, one can have several observations.}

First, we focus on the comparison of magnitude-based methods. When compared with advanced magnitude-based methods, our proposed MEB-Net outperforms other baselines consistently. Taking the performance under the seen speaker case as an example, MEB-Net provides average 0.19 PESQ, 2.15\% ESTOI and 1.77\rm{dB} SDR score improvements than BiCRN on Babble noise, while average 0.21 PESQ, 1.50\% ESTOI and 1.30\rm{dB} SDR score improvements are provided on factory1 noise. It fully demonstrates the superiority of the proposed AIA transformer in speech quality and intelligibility.  

Second, {when focusing on complex-spectrum-based methods, we can observe that most complex-spectrum-based baselines consistently surpass magnitude-based approaches.} Take BiCRN and BiGCRN as an example, we find that average 0.30 PESQ, 5.68\% ESTOI and 2.84\rm{dB} SDR score improvements are provided in the unseen speakers on babble noise. It indicates the importance of phase recovery to improve speech quality and intelligibility. Meanwhile, our proposed CPB-Net also outperforms most single-stage complex-spectrum-based methods, which all aim at optimizing magnitude and phase in a single stage. For example, in the seen speaker condition, CPB-Net provides average 0.20 PESQ, 4.27\% ESTOI and 1.01\rm{dB} SDR score improvements than BiDCCRN on Babble noise, while average 0.18 PESQ, 3.09\% ESTOI and 1.11\rm{dB} SDR score improvements are provided on factory1 noise. A similar tendency is also observed in the unseen speaker case.

Third, when the magnitude estimation and phase recovery are decoupled into two stages, consistently better performance can be obtained than single-stage complex spectrum estimation methods. For example, when babble noise is given in the seen speaker, CTS-Net achieves average 0.14, 3.01\%, and 1.30\rm{dB} score improvements over BiGCRN in terms of PESQ, ESTOI, and SDR, respectively. For factory1 noise in the seen speaker, the improvements are 0.08 PESQ, 2.59\% ESTOI and 1.31\rm{dB} SDR, respectively. Under the unseen speaker condition, similar improvements can be also obtained. This indicates the notable advantage of decoupling-based methods over single-stage methods in the complex-valued spectral domain. Then, one can see that when the magnitude and complex spectral details are optimized in parallel instead of in the cascaded pipeline, DBT-Net dramatically surpasses CTS-Net by a considerable margin. For instance, for babble noise condition in the unseen speaker case, DBT-Net achieves average 0.19, 3.31\%, and 1.46\rm{dB} score improvements over CTS-Net in terms of PESQ, ESTOI, and SDR, respectively. This indicates the merit and effectiveness of the proposed dual-branch pipeline in improving speech quality and intelligibility in the complex-valued spectral domain.

Finally, we compare our proposed method with the advanced time-domain systems. {From Tables~{\ref{tbl:seen-babble-factory1-objective}} and~{\ref{tbl:unseen-babble-factory1-objective}}, one can find that in both seen and unseen speaker cases, DBT-Net considerably outperforms AECNN in terms of all metrics,} e.g., around 0.49, 9.67\%, and 2.80\rm{dB} average improvements in PESQ, ESTOI, and SDR are observed for babble and factory1 noises in both seen and unseen speaker cases. This demonstrates that our proposed approach in the complex-spectrum domain enjoys significant performance superiority over the previous time-domain system. {Additionally, when compared with other advanced dual-path time-domain methods, one can get that DBT-Net consistently achieves better performance in terms of all metrics, indicating the superiority of the proposed dual-branch strategy and the attention-in-attention transformer-based network. For example, for the factory1 noise in the seen speaker case, DBT-Net achieves average 0.14, 4.45\%, and 0.82\rm{dB} score improvements over TSTNN in terms of PESQ, ESTOI, and SDR, respectively.}

\begin{figure}[t]
	\centering
	\centerline{\includegraphics[width=1\columnwidth]{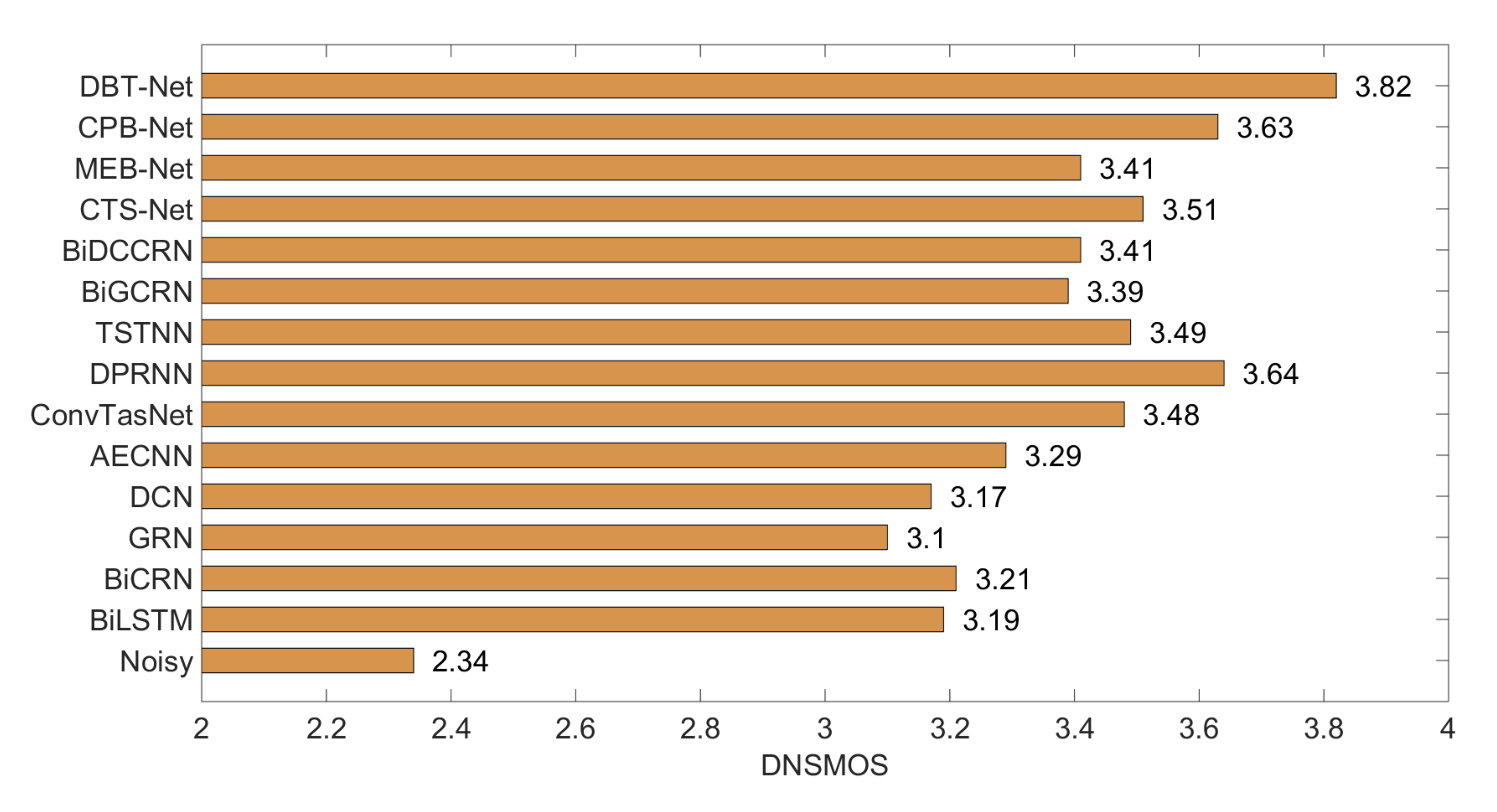}}
	\vspace{-0.4cm}
	\caption{{DNSMOS scores of different SE systems.} }
	\vspace{-0.3cm}
	\label{fig:dnsmos}	
\end{figure}

\begin{figure*}[ht]
	\centering
	\centerline{\includegraphics[width=1.95\columnwidth]{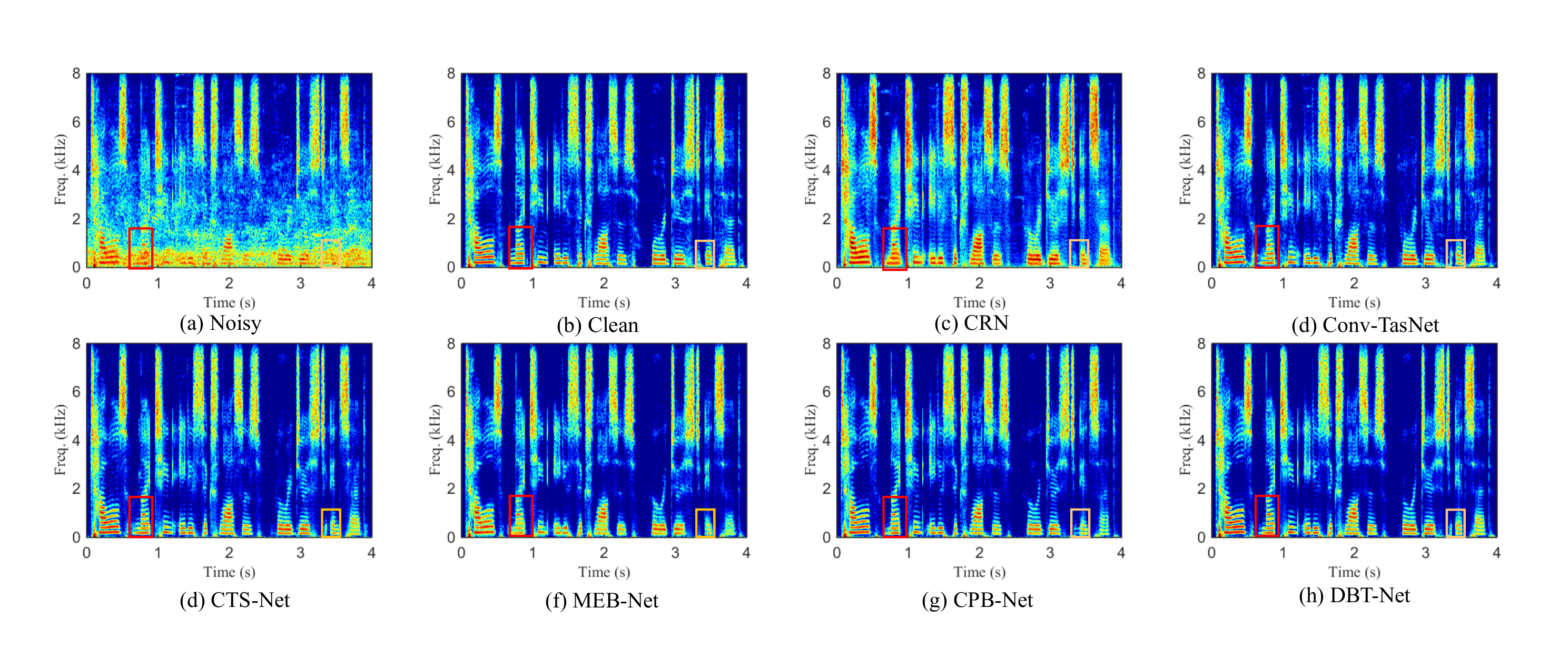}}
	\vspace{-0.6cm}
	\caption{Visualization of spectrograms of enhanced speech signals obtained from different models. (a) Noisy speech, PESQ = 1.56. (b) Clean speech, PESQ = 4.50. (c) Enhanced speech by CRN, PESQ = 2.28. (d) Enhanced speech by ConvTasNet, PESQ = 2.71. (e) Enhanced speech by CTS-Net, PESQ = 2.76. (f) Enhanced speech by MEB-Net, PESQ = 2.32. (g) Enhanced speech by CPB-Net, PESQ = 2.61. (h) Enhanced speech by DBT-Net, PESQ = 2.93. }
	\label{fig:visualization}
\end{figure*}

Besides, the perceptual evaluation on DNSMOS, a non-intrusive perceptual metric, is also provided in Fig.~{\ref{fig:dnsmos}}. One can find that the proposed approach outperforms all the advanced baselines by a significant margin, which also demonstrates the superiority of DBT-Net in improving subjective perceptual speech quality. Moreover, an example of spectrograms of clean utterance, noisy utterance the enhanced utterances by the CRN, ConvTasNet, CTS-Net, MEB-NET, CPB-Net and DBT-Net are presented in Fig.~{\ref{fig:visualization}} (a)-(h). It is obvious that DBT-NET outperforms other baselines in terms of restoring spectral details and suppressing background noise. Focusing on the red and orange boxes in Fig.~{\ref{fig:visualization}} (f) and (g), one can see that MEB-Net suppresses more background noise than CPB-Net, while CPB-Net can estimate more lost fine-gained spectral details. By merging these two branches, DBT-Net can encourage strengths and bypass the weakness of each branch from a complementary aspect, resulting in better spectrum estimating than the single-branch paradigm, as illustrated in Fig.~{\ref{fig:visualization}} (h).

\vspace{-0.3cm}
\subsection{Performance comparison with baselines using VoiceBank + DEMAND dataset\label{Section51}}

\renewcommand\arraystretch{1.1}
\begin{table}[t]
	\caption{Comparison with other state-of-the-art methods including time and T-F domain methods. “$-$" denotes that the result is not provided in the original paper.}
	\label{tbl:VB-results}
	\centering
	\resizebox{0.5\textwidth}{!}{
		\begin{tabular}{l|l|l|ccccc}
			\hline
			\multicolumn{1}{l|}{\textbf{Methods}}&\textbf{Year} &\multicolumn{1}{l|}{\textbf{Param.}} & \textbf{PESQ} & \textbf{STOI(\%)}  & \textbf{CSIG} & \textbf{CBAK} & \textbf{COVL}   \\ \hline
			\multicolumn{1}{l|}{Noisy} & {\makecell[c]{--}}  & \multicolumn{1}{l|}{\makecell[c]{--}} & 1.97  & 92.1 & 3.35 & 2.44 & 2.63 \\ \hline
			\multicolumn{8}{c}{\textbf{SOTA time and T-F Domain approaches}} \\ \hline
			\multicolumn{1}{l|}{SEGAN~{\cite{pascual2017segan}}} &2017  &\multicolumn{1}{l|}{43.2 M} & 2.16 & 92.5 & 3.48 & 2.94 & 2.80 \\ 
			\multicolumn{1}{l|}{MMSEGAN~{\cite{soni2018time}}} &2018  &\multicolumn{1}{l|} {\makecell[c]{--}}& 2.53 & 93.0 & 3.80 & 3.12 & 3.14  \\
			\multicolumn{1}{l|}{SERGAN~{\cite{baby2019sergan}}} &2019  &\multicolumn{1}{l|}{43.2 M} & 2.16 & 92.5 & 3.48 & 2.94 & 2.80 \\  	
			\multicolumn{1}{l|}{MetricGAN~{\cite{fu2019metricgan}}} &2019  &\multicolumn{1}{l|}{1.86 M}& 2.86 & {\makecell[c]{--}} & 3.99 & 3.18 & 3.42 \\ 
			\multicolumn{1}{l|}{CRGAN~{\cite{zhang2020loss}}}&2020 &\multicolumn{1}{l|}{\makecell[c]{--}} & 2.92 & 94.0 & 4.16 & 3.24 & 3.54 \\ 
			\multicolumn{1}{l|}{DCCRN~{\cite{hu2020dccrn}}}&2020  &\multicolumn{1}{l|}{3.70 M} & 2.68 & 93.7 & 3.88 & 3.18 & 3.27  \\ 		
			\multicolumn{1}{l|}{RDL-Net~{\cite{nikzad2020deep}}} &2020 & \multicolumn{1}{l|}{3.91 M} & 3.02 & {\makecell[c]{93.8}}& 4.38 & 3.43 & 3.72 \\ 
			
			\multicolumn{1}{l|}{PHASEN~{\cite{yin2020phasen}}}&2020  &\multicolumn{1}{l|}{\makecell[c]{--}}& 2.99 & {\makecell[c]{--}} & 4.21 & 3.55 & 3.62  \\
			\multicolumn{1}{l|}{MHSA-SPK~{\cite{koizumi2020speech}}}&2020  &\multicolumn{1}{l|}{\makecell[c]{--}}& 2.99 & {\makecell[c]{--}} & 4.15 & 3.42 & 3.53  \\ 
			\multicolumn{1}{l|}{T-GSA~{\cite{kim2020t}}} &2020 &\multicolumn{1}{l|}{\makecell[c]{--}} & 3.06 & 93.7 & 4.18 & 3.59 & 3.62\\ 
			\multicolumn{1}{l|}{TSTNN~{\cite{wang2021tstnn}}} &2021  &\multicolumn{1}{l|}{0.92 M}& 2.96 & 95.0 & 4.17 & 3.53 & 3.49 \\ 
			\multicolumn{1}{l|}{DEMUCS ~{\cite{defossez2020real}}}&2021  &\multicolumn{1}{l|}{128 M}& 3.07 & 95.0 & 4.31 & 3.40 & 3.63 \\ 
			\multicolumn{1}{l|}{CTS-Net~{\cite{li2021two}}} &2021 &\multicolumn{1}{l|}{4.35 M}& 2.92 & {\makecell[c]{--}} & 4.25 & 3.46 & 3.59 \\ 
			\multicolumn{1}{l|}{GaGNet~{\cite{li2021glance}}} &2021 &\multicolumn{1}{l|}{5.94 M}& 2.94 & 94.7 & 4.26 & 3.45 & 3.59 \\ 
			\multicolumn{1}{l|}{MetricGAN+~{\cite{fu2021metricgan+}}} &2021 &\multicolumn{1}{l|}{\makecell[c]{--}}& 3.15 & {\makecell[c]{--}} & 4.14 & 3.16 & 3.64 \\ 
			\multicolumn{1}{l|}{SE-Conformer~{\cite{eesung2021se}}} &2021  &\multicolumn{1}{l|}{\makecell[c]{--}}& 3.13 & 95 & 4.45 & 3.55 & 3.82 \\
			
			\hline
			
			\multicolumn{8}{c}{\textbf{Proposed approaches}} \\ \hline
			\multicolumn{1}{l|}{MEB-Net} &2021 &\multicolumn{1}{l|}{0.90 M} & 3.11 & 94.9 & 4.45 & 3.60 & 3.79 \\ 
			\multicolumn{1}{l|}{CPB-Net} &2021 &\multicolumn{1}{l|}{1.18 M} & 3.15 & 94.7 & 4.48 & 3.54 & 3.81\\ 
			\multicolumn{1}{l|}{DBT-Net} &2021  &\multicolumn{1}{l|}{2.91 M} & \textbf{3.30} &  \textbf{95.7} & \textbf{4.59} & \textbf{3.75} & \textbf{3.92} \\ \hline
		\end{tabular}
	}
\end{table}

Besides the WSJ0-SI84 corpus, we also conduct the experiments on another public benchmark, \emph{i.e.}, VoiceBank + DEMAND, to further validate the superiority of the proposed approach with other SOTA baselines, whose results are presented in Table~{\ref{tbl:VB-results}}. {From the results in Table~{\ref{tbl:VB-results}}, one can have the following observations.} First, when only the magnitude estimation branch (MEB-Net) or the complex spectral mapping branch (CPB-Net) is adopted, the proposed frameworks achieve competitive performance compared with most existing advanced single-branch based baselines in the magnitude or complex spectrum domain. For example, from RDL-Net to MEB-Net, average 0.09, 1.1\%, 0.07, 0.17 and 0.07 improvements are achieved in terms of PESQ, STOI, CSIG, CBAK and COVL, respectively. 
Similarly, CPB-Net provides average 0.09 PESQ, 1.0\%STOI, 0.30 CSIG, and 0.19 COVL improvements over T-GSA. This verifies the effectiveness of the proposed attention-in-attention transformer-based network in improving speech quality. {In addition, when only the single-branch topology is adopted, one can observe that CPB-Net yields better performance in PESQ, CSIG and COVL than MEB-Net, while MEB-Net achieves a higher score in CBAK. This indicates that MEB-Net can better eliminate noise, while CPB-Net conducts better speech overall quality. Second, by simultaneously adopting two branches in parallel, DBT-Net yields significant improvements in terms of all metrics than the single-branch methods. This verifies that the proposed dual-branch method can collaboratively facilitate the complex spectrum recovery from the complementary perspective. Compared with other existing single-stage and decoupling-style SOTA methods, DBT-Net achieves consistently better speech performance.} For example, from the previous decoupling-style approach, \emph{i.e.}, GaGNet, to DBT-Net, average 0.36, 0.33, 0.30 and 0.33 improvements can be observed in terms of PESQ, CSIG, CBAK and COVL respectively. Third, compared with previous SOTA time-domain baselines, the proposed DBT-Net also achieves better performance in all objective metrics. For example, DBT-Net provides average 0.17, 0.14, 0.20, and 0.10 improvements over SE-Conformer in terms of PESQ, CSIG, CBAK and COVL, respectively.

Additionally, we also provide a comparison of the number of parameters between our methods and some reported SOTA methods. The proposed method has a relatively low parameter burden, \emph{i.e.}, 2.91 M, when compared with the most advanced time-domain and T-F domain methods.

\section{Conclusions\label{Section6}}
\label{Sec5}

In this paper, we propose a dual-branch transformer-based framework to federatively facilitate clean spectrum estimation from the complementary perspective. Specifically, a magnitude spectrum estimation branch (MEB) is designed to coarsely filter out the dominant noise components in the magnitude domain, while the residual spectral details are derived by a complex spectrum purification branch (CPB) in parallel. 
To leverage the information exchange between two branches, the interaction block is proposed to guide the sequence modeling by the information learned from the other branch. Within each branch, we introduce a novel attention-in-attention transformer-based (AIAT) module between a densely encoder-decoder architecture for contextual information modeling, which aims to strengthen long-term spectro-temporal dependencies and aggregate global hierarchical intermediate information. Experimental results on two public datasets, namely WSJ0-SI84 + DNS Challenge and VoiceBank + DEMAND, demonstrate that the proposed method achieves state-of-the-art performance over previous advanced approaches remarkably in various objective and subjective metrics.

\bibliographystyle{IEEEbib}
\bibliography{myrefs} 

\end{document}